\documentclass[a4paper,11pt,twoside]{article}

\usepackage{geometry} 
\usepackage{amsmath}
\usepackage{amssymb}
\usepackage{color}
\usepackage{graphicx}

\usepackage{axodraw}
\usepackage{subfig}

\newcommand{\Sb}{\overline{16}}
\newcommand{\Sp}{16^\prime}
\newcommand{\Sbp}{\overline{16}^\prime}
\newcommand{\Spp}{16^{\prime \prime}}
\newcommand{\Sbpp}{\overline{16}^{\prime \prime}}

\newcommand{\Sib}{\overline{S}}
\newcommand{\Sip}{S^\prime}
\newcommand{\Sibp}{\overline{S}^\prime}
\newcommand{\Sipp}{S^{\prime \prime}}
\newcommand{\Sibpp}{\overline{S}^{\prime \prime}}

\newcommand{\Tp}{10^\prime}
\newcommand{\Up}{U^\prime}
\newcommand{\MS}{M_{54}}
\newcommand{\MA}{M_{45}}
\newcommand{\Mhun}{M_{120}}
\newcommand{\MT}{ M_{10}}
\newcommand{\la}[1]{\lambda_{#1} \,}

\newcommand{\TeV}{\,\mathrm{TeV}}
\newcommand{\GeV}{\,\mathrm{GeV}}
\newcommand{\MeV}{\,\mathrm{MeV}}

\newcommand{\mgut}{{M_\text{GUT}}}

\newcommand{\fracwithdelims}[4]{\left#1 \frac{#3}{#4} \right#2}
\newcommand{\ord}[1]{\mathcal{O}\left( #1 \right)}

\newcommand{\vev}[1]{\left\langle #1\right\rangle}
\newcommand{\VeV}[2]{#1\langle #2 #1\rangle} 

\newcommand{\Fig}[1]{Fig.~\ref{fig:#1}}

\newcommand{\Eq}[1]{Eq.~(\ref{eq:#1})}
\newcommand{\eq}[1]{eq.~(\ref{eq:#1})}

\newcommand{\eqs}[1]{eqs.~(\ref{eq:#1})}

\newcommand{\interskip}{\medskip}
 
\newcommand{\nohyphens}%
        {\hyphenpenalty=10000\exhyphenpenalty=10000\relax}
\newcommand{\omi}[1]{}

\DeclareMathOperator{\re}{Re}
\DeclareMathOperator{\tr}{Tr}
\DeclareMathOperator{\str}{Str}

\newcommand{\GSM}{G_\text{SM}}

\newlength{\myem}
\newcommand{\sep}[1]{#1}
\newcounter{mysubequation}[equation]
\renewcommand{\themysubequation}{\alph{mysubequation}}
\newcommand{\mytag}{\stepcounter{mysubequation}%
\tag{\theequation\protect\sep{\themysubequation}}}
\newcommand{\globallabel}[1]{\refstepcounter{equation}\label{#1}}

\newcommand{\SISSA}{SISSA/ISAS and INFN, I--34151 Trieste, Italy}

\newcommand{\titletext}{General Aspects of Tree Level Gauge Mediation} 
\newcommand{\authortext}{\large Marco Nardecchia, Andrea Romanino, and Robert Ziegler
\medskip\\\em\normalsize \SISSA}
\newcommand{\abstracttext}{Tree level gauge mediation (TGM) may be considered as the simplest way to communicate supersymmetry breaking: through the tree level renormalizable exchange of heavy gauge messengers. We study its general structure, in particular the general form of tree level sfermion masses and of one loop, but enhanced, gaugino masses. This allows us to set up general guidelines for model building and to identify the hypotheses underlying the phenomenological predictions. In the context of models based on the ``minimal'' gauge group SO(10), we show that only two ``pure'' embeddings of the MSSM fields are possible using $d< 120$ representations, each of them leading to specific predictions for the ratios of family universal sfermion masses at the GUT scale,  $m^2_{\overline{5}} = 2 m^2_{10}$ or  $m^2_{\overline{5}} = (3/4) m^2_{10}$ (in SU(5) notation). These ratios are determined by group factors and are peculiar enough to make this scheme testable at the LHC. We also discuss three possible approaches to the $\mu$-problem, one of them distinctive of TGM. }

\title{
\normalsize
\hspace*{\fill}
\begin{tabular}[t]{l}
\end{tabular}
\vspace{3\baselineskip}\\\Large\bfseries\titletext\bigskip}
\author{\begin{minipage}[t]{0.8\textwidth}
\normalsize\centering\authortext
\end{minipage}}
\date{}

\begin{document}

\bigskip
\maketitle
\begin{abstract}\normalsize\noindent
\abstracttext
\end{abstract}\normalsize\vspace{\baselineskip}



\section{Introduction}

In the long theoretical preparation for the LHC, a wide spectrum of options for the new physics at the TeV scale has been considered. A major role has been played by supersymmetric models. Several schemes have been investigated in which supersymmetry is broken in a hidden sector with no renormalizable interactions with the observable sector and is communicated to the latter by a variety of mechanisms~\cite{GravMSB,GMSB,AMSB,GauginoMSB}. Below the scale $M$ at which supersymmetry breaking is communicated, sfermion masses are typically described by the model-independent effective lagrangian operator
\begin{equation}
\label{eq:sfermioneffective}
\int d\theta^2 d\overline{\theta}^2 \,
\frac{Z^{\dagger}Z Q^{\dagger} Q}{M^2} ,
\end{equation}
where $Z$ is a Standard Model (SM) singlet hidden chiral superfield whose $F$-term vev breaks supersymmetry, $\vev{Z} = F \theta^2$, and $Q$ is a generic light, observable chiral superfield, for example an MSSM one. Such an effective, supersymmetric description holds if $F\ll M^2$. Different models are characterized by different origins for the above operator. 

Surprisingly enough, a simple and attractive possibility has been neglected in the almost three decades of  phenomenological studies of supersymmetry: the possibility that the operator in \eq{sfermioneffective} arises from the renormalizable, tree level exchange of heavy vector superfields, as in \Fig{diagram}. This is the communication mechanism that we call tree level gauge mediation (TGM)~\cite{NRZ}. Besides its simplicity (compare for example with the cumbersome set of two loop diagrams generating the operator in \eq{sfermioneffective} in ordinary, loop gauge mediation), this framework is also motivated by the necessary presence of superheavy vector fields in Grand Unified Theories (GUTs). On the verge of the LHC era, we believe it is worth filling this lacuna and spell out the consequences of TGM, also in the light of its peculiar predictions. 

One may wonder how such a simple possibility could have been missed. The reason might be that well known arguments seem to prevent it. The main obstacle is represented by the supertrace formula~\cite{MassSumRule} and its consequences. In the context of the tree level, renormalizable spontaneously broken supersymmetric theory underlying \Fig{diagram}, we must have
\begin{equation}
\label{eq:str}
\str\mathcal{M}^2 = g D_a \tr(T_a) ,
\end{equation}
for the supertrace of the squared masses of the fields in the model. \Eq{str} holds separately for each set of conserved quantum numbers~\cite{GeorgiDimopoulos}. If the action of the gauge generators on the full set of chiral superfields, $T_a$, is traceless, as in the case we are going to consider, the supertrace vanishes. This represents a potential phenomenological problem. Still, TGM leads to a viable spectrum, as we will see. While the explicit construction in the next Sections is all we need to get to our results, we find useful, in this introduction, to review what the potential problem is and illustrate how tree level gauge mediation solves it. 

We can see the potential problem at two different levels. First, \eq{str} holds in particular when applied to all fields with the quantum numbers of the SM fermions. Let us consider then the case of the MSSM. In this case the fields with the quantum numbers of the SM fermions are the SM fermions themselves, $f$, and their supersymmetric partners, the sfermions $\tilde f$. From $\str \mathcal{M}^2_\text{$f,\tilde f$} = 0$ we then conclude that the sum of the squared masses of fermions and sfermions should coincide. This is in clear contradiction with the experimental bounds on the sfermion masses, giving $\str \mathcal{M}^2_\text{$f,\tilde f$} > 0$. This is however far from being the end of the story, as any realistic complete theory of supersymmetry breaking is likely to involve additional fields on top of the MSSM ones. This is the case of our TGM framework, where the positive contribution to the supertrace from the MSSM fermions and sfermions is compensated by an opposite contribution from extra fields with quantum numbers within the ones of the SM fermions, $\str \mathcal{M}^2_\text{extra} < 0$, so that $\str \mathcal{M}^2_\text{$f,\tilde f$} + \str \mathcal{M}^2_\text{extra} = 0$, in agreement with the supertrace formula. The extra chiral superfields will get heavy supersymmetric mass terms. Their negative contribution to the supertrace is due to the fact that their scalar components get negative $\ord{\text{TeV}}$ soft masses, which however represent only negligible corrections to their much larger, positive supersymmetric mass term. As we will see, this can be obtained without ad hoc model building efforts. 

The supertrace formula has stronger implications than the ones outlined above, which should also be addressed. Let us consider the fields with the $\text{SU(3)}_c\times\text{U(1)}_\text{em}$ quantum numbers of the $d$ quarks only. The latter set will contain at least the three down-type SM quarks and their scalar partners and possibly the extra fields we need to compensate the supertrace formula. Let us project the supertrace formula in the flavor space of down-type fields along the direction corresponding to the lightest $d$-quark mass eigenstate, the down quark. Note that when restricted to a given set of quantum numbers, the trace on the right-hand side of \eq{str} can be non-vanishing. Assume now that the only U(1) factor in the gauge group is the SM hypercharge, $\text{U(1)}_Y$. We then obtain~\cite{GeorgiDimopoulos,W3}
\begin{equation}
\label{eq:constraint}
m^2_{\tilde d} \leq m^2_d -\frac{1}{3} g' D_Y,
\end{equation}
where $\tilde d$ is the lightest $d$-sfermion mass eigenstate, $m_d$ is the down quark mass, $m_d \sim 
5\MeV$, $g'$ is the hypercharge gauge coupling and $D_Y$ the hypercharge $D$-term. \Eq{constraint} represents a serious phenomenological problem, even in the presence of the extra fields invoked above. If $D_Y=0$, in fact, \eq{constraint} would force a down sfermion mass to be smaller than about $5\MeV$, in contrast with the lowest experimental limits of a few hundreds GeV. If $D_Y > 0$, the constraint would be even stronger. If $D_Y < 0$, the constraint would be loosened, but one could repeat the argument for the up quarks and squarks. For which the $D_Y$ contribution to the relation analogous to \eq{constraint} would have opposite sign, leading to an even stronger bound for the lightest up squark. In order to bypass this problem, an extra U(1) factor, giving the same sign on both down and up fields, is needed. Such an extra U(1) factor is present ``by definition'' in the TGM scheme. It is the U(1) factor associated to the heavy vector exchange in \Fig{diagram} (as $Z$ is a SM singlet, the heavy vector must also be a SM singlet). We therefore have all the ingredients needed to overcome the potential problem set by the supertrace formula. As we will see, those ingredients naturally combine in phenomenologically viable schemes. 

Another potential problem is represented by the fact that gaugino masses arise at the loop level and are therefore potentially suppressed with respect to the sfermion masses by a large loop factor, thus pushing the sfermions out of the reach of the LHC and introducing a significant fine-tuning in the determination of the Higgs mass. We will list in Section~\ref{sec:gauchi} a number of gaugino mass enhancement factors that can compensate fully or partially that loop factor. 

A minimal model of tree-level gauge mediation has been presented in~\cite{NRZ}, solving the supersymmetric flavor problem and predicting the ratio of different sfermion masses to be different from mSugra and other schemes. In this paper we would like to take a broader point of view and study the general implementation of TGM. This will allow to establish the general properties of TGM, to set up the guidelines for model building and to identify what are the hypotheses under which the peculiar predictions on soft masses of the minimal model hold. Moreover, we would like to present a few new approaches to the $\mu$-problem, both in the context of well known (Giudice-Masiero, NMSSM) and new solutions.  In particular, in Section~\ref{sec:tree} we will discuss what are the conditions under which heavy vector superfields can act as tree-level messengers of supersymmetry breaking and obtain a general expression for the tree level contribution to the supersymmetry breaking lagrangian, in particular to the sfermion soft masses. In Section~\ref{sec:oneloop}, we will consider the one-loop contributions to soft masses, concentrating mostly on gaugino masses and the enhancement factors compensating their loop suppression. In Section~\ref{sec:guidelines} we consider the possibility to obtain a phenomenologically viable model from the general formalism previously introduced. We will see that clear model building guidelines emerge, leading to peculiar predictions for the pattern of MSSM sfermion masses and we will identify the assumptions underlying such predictions. In Section~\ref{sec:mu}, we will discuss a few new approaches to the $\mu$-problem, before summarizing in Section~\ref{sec:conclusions}. The paper also contains two Appendixes. In Appendix A, we outline the procedure to integrate out vector superfields and address a few minor issues, such as the generalization to the non-abelian case and the role of gauge invariance in a consistent supersymmetric generalization of the expansion in the number of derivatives. In Appendix B, we provide an example of a superpotential achieving supersymmetry breaking, SO(10) breaking to the SM, ensuring that only the MSSM fields survive at lower energy (in particular providing doublet-triplet splitting) and solving the $\mu$-problem. Such a superpotential is not aimed at being simple or realistic, but it represents a useful existence proof. 

\section{Tree level soft terms}
\label{sec:tree}

In this Section we discuss the conditions under which heavy vector superfields can act as tree-level messengers of supersymmetry breaking in the context of a generic, renormalizable, $N=1$ globally supersymmetric gauge theory in four dimensions. Then we recover the general expression for the tree level contribution to the sfermion soft masses. We discuss their origin both in the context of the full, renormalizable theory, and in an effective theory approach.   

We start from a lagrangian described by a canonical K\"ahler $K = \Phi^\dagger e^{2gV}\Phi$ and gauge kinetic function and by a generic superpotential $W(\Phi)$ function of the chiral superfields $\Phi\equiv (\Phi_1\ldots\Phi_n)$, with no Fayet-Iliopoulos term. We follow the conventions in~\cite{WessBagger}. We will denote by $\phi_i$, $\psi_i$, $F_i$ the scalar, spinor, and auxiliary component of $\Phi_i$ and by $v^\mu_a$, $\lambda_a$, $D_a$ the vector, spinor, and auxiliary component of $V_a$. The gauge group $G$ (assumed for simplicity to be simple with a single gauge coupling $g$) is broken by the scalar component vev $\phi_0 = \vev{\phi}$ to the subgroup $H$ at a scale $M_V \sim g |\phi_0|\gg M_Z$, at which the theory is approximately supersymmetric. In the phenomenological applications we have in mind, $H$ contains the SM gauge group $\GSM$, $G$ is a grand-unified group (for example SO(10) or $E_6$), and the breaking scale is of the order of the GUT scale. Correspondingly, the vector superfields split into light and heavy ones, associated to the orthonormalized generators $T^l_a$ and $T^h_b$ respectively: $V = V_a^l T_a^l + V_b^h T_b^h$, $a=1\ldots N_l$, $b=1\ldots N_h$. 

The heavy vector superfields acquire a squared mass matrix given by 
\begin{equation}
\label{eq:MMV0}
(M^2_{V0})_{ab}   = g^2 \phi_0^\dagger \{ T_a^h, T_b^h \} \phi_0. 
\end{equation}
We choose the basis of heavy generators $T^h_a$ in such a way that the above mass matrix is diagonal,
\begin{equation}
\label{eq:MMV0diag}
(M^2_{V0})_{ab}   = M^2_{V_a} \delta_{ab} .
\end{equation}
The heavy vector superfields become massive by eating up a corresponding number of Goldstone chiral superfields. It is then convenient to split the chiral superfields as follows
\begin{equation}
\label{eq:goldstones}
\Phi = \phi_0+\Phi' + \Phi^G, \quad \Phi^G = \sqrt{2}\, g \frac{\Phi^\text{G}_a}{M_{V_a}} T^h_a \phi_0 ,
\quad \Phi' = \Phi'_i b_i,
\end{equation}
where $\Phi^G_a$, $a=1\ldots N_h$ are the Goldstone superfields associated to the generators $T^h_a$ and $b_i = (b^i_1\ldots b^i_n)$, $i=1\ldots n-N_h$ is an orthonormal basis in the space of the ``physical'' chiral fields $\Phi'$, $b^\dagger_i T_a \phi_0 = 0$. In the supersymmetric limit, $\phi_0$ is orthogonal to $\Phi^G$ and $\Phi^G$ does not mix with the physical superfields. The physical components of the massive vector superfield $V_a$ are $v^\mu_a$, $\lambda_a$, $\psi^G_a$, $\re(\phi^G_a)/\sqrt{2}$, all with mass $M_{V_a}$. The imaginary part of $\phi^G_a$, the Goldstone boson, becomes as usual the longitudinal component of the massive gauge boson $v^\mu_a$ and the spinors $\psi^G_a$ and $\lambda_a$ pair up in a Dirac mass term. This spectrum can be split by supersymmetry breaking corrections, as we will see in Section~\ref{sec:gauginovector}. 

As for the physical chiral superfields $\Phi'_i$, their supersymmetric mass matrix is given by 
\begin{equation}
\label{eq:MC0}
M^0_{ij} = \frac{\partial^2 W}{\partial \Phi'_i \partial \Phi'_j}(\phi_0) .
\end{equation}
Again, we choose the basis $b_i$ in such a way that the above mass matrix is diagonal and positive,
\begin{equation}
\label{eq:MC0diag}
M^0_{ij} = M_i \delta_{ij}, \quad M_i \geq 0 .
\end{equation}
The scalar and fermion components of $\Phi'$ can be split by supersymmetry breaking corrections, which can also induce a mixing with the scalar and fermion components of the heavy vector superfields. 

Supersymmetry is supposed to be broken at a much lower scale than $M_V$, where some of the fields $\Phi'$ get an $F$-term, $\vev{\Phi'} = F_0\theta^2$, $M^2_Z\ll |F_0|\ll M^2_V$. As a consequence,  $\phi_0$ satisfies with good approximation the $F$-term and $D$-term conditions at the scale $M_V$, $\partial_i W(\phi_0) = 0+\ord{|F_0|}$ and $\phi_0^\dagger T_a \phi^{\phantom{\dagger}}_0 = 0 + \ord{|F_0/M_V|^2}$ (see \eq{Dh}) for each $i,a$. 

The $F$-terms induce a non vanishing vev for the $D$-terms $D^h_a$ of the heavy vector superfields. The stationary condition for the scalar potential $V$,  $ \partial V/ \partial\phi_i   =0 $, together with the gauge invariance of the superpotential give
\begin{equation}
\label{eq:Dh}
\VeV{\big}{D^h_a} = -2g \frac{F_0^\dagger T^h_a F^{\phantom{\dagger}}_0}{M^2_{V_a}},
\end{equation}
with the light $D$-terms still vanishing. Clearly, only generators $T^h_a$ that are singlets under the unbroken group $H$ can contribute to such $D$-term vevs. Note also the condition 
\begin{equation}
\label{eq:Fgauge}
F^\dagger_0 T_a \phi_0 = 0 ,
\end{equation}
which implies that the Goldstone superfields $\Phi^G$ do not get $F$-term vevs (the $D$-term condition implies that in the supersymmetric limit they do not get scalar vev either). The latter relation also follows from the gauge invariance of the superpotential. In turn, the $D$-terms above give rise to tree level soft masses for the scalar components $\phi'_i$ of the chiral superfields $\Phi'_i$
\begin{gather}
\label{eq:Dhsfermion}
V \supset \frac{1}{2} D^2 \supset -g\, \phi'^\dagger T^h_a \phi' \VeV{\big}{D^h_a} = (\tilde m^2_{ij})_D \phi'^\dagger_i \phi'_j \\
\label{eq:mmD}
(\tilde m^2_{ij})_D = 2 g^2 (T^h_a)_{ij} \frac{F_0^\dagger T^h_a F^{\phantom{\dagger}}_0}{M^2_{V_a}}  ,
\end{gather}
provided that both $F_0$ and the scalars $\phi'$ are charged under the (broken) gauge interaction associated to $T^h_a$ and provided that $T^h_a$ is a singlet under $H$. 

The complete list of tree level soft terms obtained from the gauge dynamics can be more conveniently recovered in the effective theory below $M_V$. Before discussing it, let us observe that this theory must necessarily satisfy the supertrace formula $\text{Str}(\mathcal{M}^2)=0$. In the case of the soft terms in \eq{mmD}, this simply follows here from $\tr{T^h_a} = 0$. In particular, the tracelessness condition implies that positive soft masses are accompanied by negative ones in \eq{Dhsfermion}. This is a potential phenomenological problem, which has long been considered as an obstacle to models in which supersymmetry breaking terms are generated, as here, at the tree, renormalizable level.  However, it has recently been shown~\cite{NRZ} that such a potential problem can be easily solved by adding a large positive supersymmetric mass term to the chiral superfields whose tachyonic nature would be problematic. 

\interskip

As mentioned, the generation of the sfermion masses can be conveniently seen in the effective theory below $M_V$, where the heavy vector and the Goldstone chiral superfields have been integrated out. In this theory, the chiral degrees of freedom are the $\Phi'$. The gauge group is $H$ and it is unbroken (we neglect electroweak symmetry breaking). As a consequence, there is no $D$-term contribution to supersymmetry breaking. The scalar masses arise in this context from $F$-terms vevs through an effective K\"ahler operator, as we will see in a moment. 

\begin{figure}
\begin{center}
\begin{picture}(400,130)(-80,-20)
\SetWidth{1.5}
\Photon(-60,50)(-10,50){4}{6}
\SetWidth{1.5}
\Line(-90,75)(-60,50)
\Line(-90,25)(-60,50)
\Line(20,75)(-10,50)
\Line(20,25)(-10,50)
\SetWidth{0.5}
\Text(-95,80)[]{ $\Phi'$ }
\Text(27,80)[]{ $\Phi'^{\dagger} $ }
\Text(-30,62)[]{ $V$ }
\Text(-98, 28)[]{ $\Phi'^{\dagger}  $ }
\Text(29, 28)[]{ $\Phi'$ }
\SetWidth{2.5}
\Text(53,50)[]{ $\longrightarrow$ }
\SetWidth{1.5}
\Photon(120,50)(180,50){4}{6}
\SetWidth{1.5}
\Line(90,75)(120,50)
\Line(90,25)(120,50)
\Line(210,75)(180,50)
\Line(210,25)(180,50)
\SetWidth{0.5}
\Text(86,80)[]{ $Z$ }
\Text(83, 28)[]{ $Z^{\dagger}  $ }
\Text(150,62)[]{ $V$ }
\Text(218,80)[]{ $Q^{\dagger}$ }
\Text(220, 28)[]{ $Q $ }
\Text(250, 50)[]{ $+$ }
\SetWidth{1.5}
\Photon(300,20)(300,80){4}{6}
\SetWidth{1.5}
\Line(330,105)(300,80)
\Line(270,105)(300,80)
\Line(330,-5)(300,20)
\Line(270,-5)(300,20)
\SetWidth{0.5}
\Text(266,108)[]{ $Z$ }
\Text(264, -3)[]{ $Z^{\dagger}  $ }
\Text(313,50)[]{ $V$ }
\Text(339,108)[]{ $Q^{\dagger} $ }
\Text(340, -3)[]{ $Q$ }
\end{picture}
\end{center}
\caption{Tree level gauge mediation supergraph generating the operator in \eq{dK} when integrating out the heavy vector superfield messengers.}
\label{fig:diagram}
\end{figure}
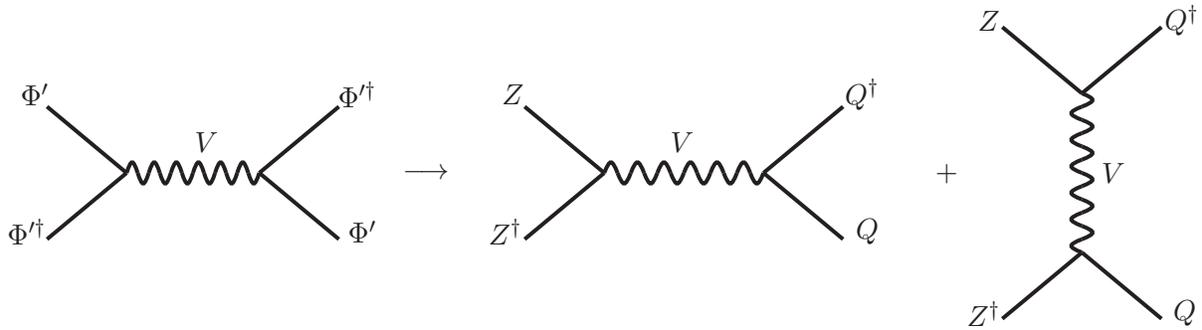

The vector superfields can be integrated out by solving the equations of motion $\partial K/\partial V^h_a = 0$~\cite{scrucca,refsthereintogether}. In Appendix~\ref{sec:eom} we illustrate the details of such a procedure in a general case, we explicitly write the resulting effective theory at the leading order, and we make a few general remarks on the approximations involved in using $\partial K/\partial V^h_a = 0$ and on the role of gauge invariance in a consistent supersymmetric generalization of the expansion in the number of derivatives~\cite{scrucca}. For the present purposes, we are only interested in the terms in the effective lagrangian relevant to (sizable) soft supersymmetry breaking. Those are the ones  following from the effective tree level contribution to the K\"ahler potential in eq.~(\ref{eq:Keff2}a):
\begin{equation}
\label{eq:dK}
\delta K^0_\text{eff} = - \frac{g^2}{M^2_{V_a}} (\Phi'^\dagger T^h_a \Phi') (\Phi'^\dagger T^h_a \Phi') ,
\end{equation}
where we remind that $\Phi'$ has no vev in its scalar component. The operator in \eq{dK} can be seen to arise from the diagram on the left-hand side in \Fig{diagram}. 

As mentioned, the only possible source of supersymmetry breaking in the effective theory are the $F$-term vevs of the chiral superfields $\Phi'$. We remind that such $F$-term vevs must belong to non-trivial representations of the full group $G$, in order to play a role in TGM. The only terms in the lagrangian containing such $F$-term vevs, at the tree level and up to second order in $F_0$, $F^\dagger_0$, and $1/M_{V_a}$, arise from the superpotential and from the operator in \eq{dK}:
\begin{multline}
\label{eq:treesoft}
- \mathcal{L}^\text{tree}_\text{soft} = -
F_{0i} 
\frac{\partial \hat W}{\partial \Phi_i}
- 2g^2 \frac{(F^\dagger_0 T^h_a \psi') (\phi'^\dagger T^h_a \psi' )}{M^2_{V_a}}
+\text{h.c.} \\  
+ 2 g^2 \frac{(F_0^\dagger T^h_a F^{\phantom{\dagger}}_0) (\phi'^\dagger T^h_a \phi')}{M^2_{V_a}}
+ 2 g^2 \frac{(\phi^\dagger T^h_a F^{\phantom{\dagger}}_0) (F^\dagger_0 T^h_a \phi')}{M^2_{V_a}} - 
F^\dagger_0 F^{\phantom{\dagger}}_0 ,
\end{multline}
where $\hat W$ is the superpotential in the effective theory,
\begin{equation}
\label{eq:What}
\hat W(\Phi') = W(\phi_0 + \Phi') \qquad \text{($\Phi^G = 0$)}.
\end{equation}
Let us consider the different terms in \eq{treesoft} in turn. The first term in the second line reproduces the contribution to the soft scalar masses in \eq{mmD}. The second term gives rise to an additional contribution, only relevant to superfields that are gauge partners of the Goldstino superfield (and have the same quantum numbers under $H$ as some of the generators of $G$)\footnote{The latter contribution can be obtained in the context of the full theory by using the unitary gauge or in Wess-Zumino gauge from the $F$-term contribution to the scalar potential using \eq{mixing} below.}. All in all, we have
\begin{equation}
\label{eq:mm}
\tilde m^2_{ij} = 2 g^2 \left[ (T^h_a)_{ij} \frac{
F^\dagger_0 T^h_a F_0}{M^2_{V_a}} + \frac{(T^h_a F_0)^*_i (T^h_a F_0)_j }{M^2_{V_a}} \right] .
\end{equation}
Note that the soft terms do not actually depend on the gauge coupling or on the normalization of the generators $T$, as $M^2_{V_a}$ is also proportional to $g^2T^2$. The second term in the first line of \eq{treesoft} is a gauge-generated Yukawa interaction with coupling $\lambda = \ord{|F_0|/M^2_V}$,  usually absent in models of supersymmetry breaking.  
From a phenomenological point of view, such tiny Yukawa couplings might play a role in neutrino physics, where they could represent naturally small Dirac neutrino Yukawa couplings~\cite{neutrinomass}.

Finally, the first term in \eq{treesoft}, has to do with the existence of a hidden sector in the effective theory. In the phenomenological applications we have in mind, the light spectrum will contain the MSSM chiral superfields, as part of a light, ``observable'' sector. The latter will be charged under the residual gauge group $H\supseteq \GSM$. On the other hand, the supersymmetry breaking superfields do not feel the residual gauge interactions. In the effective theory, therefore, the supersymmetry breaking sector is hidden from the observable sector from the point of view of gauge interactions. In order for the supersymmetry breaking sector to be hidden also from the point of view of superpotential interactions, it is sufficient to make sure that the first term in \eq{treesoft} does not induce a direct coupling between the two sectors. To be more precise, we can write the chiral superfields of the effective theory, $\Phi'$, as
\begin{equation}
\label{eq:lightheavy}
\Phi' = (Z, Q, \Phi^h) .
\end{equation}
The superfield $Z$ is the only one getting an $F$-term vev, $\vev{Z} = |F_0| \theta^2$. Its fermion component is the Goldstino and therefore $Z$ is a massless eigenstate of the mass matrix $M^0$ in \eq{MC0}. The remaining mass eigenstates are divided in two groups, the heavy ones, $\Phi^h_i$  with masses $M^h_i \gg |F_0|$, and the light, or observable, ones $Q_i$, with masses $M^Q_i \lesssim |F_0|$. In order to hide supersymmetry breaking from the observable sector also from the point of view of superpotential interactions, we require that 
\begin{equation}
\label{eq:hide}
\frac{\partial^2 \hat W}{\partial Z \partial Q_j}(Z,Q,\Phi^h=0) = 0 
\end{equation}
(at least for the renormalizable part of the superpotential). 

We can then see supersymmetry breaking as arising in a hidden sector and then communicated from the to the observable sector by the diagrams on the right-hand side of \Fig{diagram}. This can perhaps be considered as the simplest way to communicate supersymmetry breaking: through the tree level renormalizable exchange of a heavy gauge messenger. Since heavy gauge messengers at a scale not far from the Planck scale are automatically provided by grand-unified theories, this possibility is not only simple but also well motivated. The reason why it has not been pursued in the past is an apparent obstacle arising from the supertrace theorem that, as mentioned, can be easily evaded by providing heavy, supersymmetric masses to some of the superfields. Such mass terms can naturally arise in the context of grand-unified theories, as we will see. 

We end this Section with some comments on integrating out heavy chiral superfields and on the corresponding possible tree level contributions to $A$-terms and soft scalar masses. The heavy vector superfields may not be the only fields living at the scale $M_V$, as chiral superfields could have mass terms of similar size or get it after gauge symmetry breaking. Such chiral fields should also be integrated out in order to get the effective theory below the scale $M_V$. In general, we want to integrate out all the heavy chiral superfields $\Phi^h$. Since their masses $M^h_i$ are assumed to be much larger than the supersymmetry breaking scale, it will still be possible to write the effective theory in a manifestly supersymmetric way. In order to integrate them out, let us write the superpotential as
\begin{equation}
\label{eq:Wexpansion}
\hat W = -|F_0| Z + \frac{M^Q_i}{2} Q_i^2 + \frac{M^h_i}{2} (\Phi^h_i)^2 + W_3(Z,Q,\Phi^h) ,
\end{equation}
where $W_3$ is at least trilinear in its argument. The equations of motion $(\partial\hat W)/(\partial \Phi^h_i) = 0$ give
\begin{equation}
\label{eq:Phih}
\Phi^h_i = -\frac{1}{M^h_i} \frac{\partial W_3}{\partial \Phi^h_i} (Z,Q) + \ord{\frac{1}{M_h^2}} .
\end{equation}
The effective superpotential for the light fields $Z$ and $Q$ is therefore 
\begin{equation}
\label{eq:Weff}
W_{\text{eff}}(Z,Q) = \hat W(Z,Q) - \frac{1}{2M^h_i}\sum_i \left(
\frac{\partial W_3}{\partial \Phi^h_i} (Z,Q)
\right)^2 + \ord{\frac{1}{M_h^2}} .
\end{equation}
A contribution to the effective K\"ahler is also induced
\begin{equation}
\label{eq:Keffchiral}
\delta K_\Phi = \frac{1}{(M^h_i)^2} \sum_i \left|
\frac{\partial W_3}{\partial \Phi^h_i} (Z,Q)
\right|^2 + \ord{\frac{1}{M_h^3}} .
\end{equation}
The effective contributions to the superpotential and to the K\"ahler in eqs.~(\ref{eq:Weff}) and~(\ref{eq:Keffchiral}) may give rise to ``chiral-mediated'' tree-level $A$-terms and (negative) additional contributions to soft scalar masses respectively. The latter should be subleading with respect  to the (positive) vector mediated contributions in \eq{mm}, at least in the case of the MSSM sfermions. Such tree level contributions could only arise in the presence of trilinear superpotential couplings in the form $ZQ\Phi^h$. In the following we will consider the case in which such a coupling is absent,
\begin{equation}
\label{eq:pure}
\frac{\partial^3 \hat W}{\partial Z \partial Q\partial \Phi^h}(0) = 0 ,
\end{equation}
so that the chiral-mediated tree tree level contributions also vanish. This is often the case, as illustrated by the model in~\cite{NRZ}. 

\section{One loop soft terms and gaugino masses}
\label{sec:oneloop}

In this Section we consider the one loop contributions to soft masses, focusing mostly on gaugino masses and the enhancement factors compensating their loop suppression. 

Gaugino masses do not arise at the tree level. They are however generated at the one-loop level, as in standard, ``loop'' gauge mediation models. The suppression of gaugino masses by a loop factor with respect to scalar masses represents a potential phenomenological problem. Given the present experimental limits on gaugino masses, a loop factor enhancement would make the sfermions heavier than $\ord{10\TeV}$, beyond the reach of the LHC and heavy enough to introduce a serious fine-tuning problem, thus approaching the split supersymmetry regime~\cite{SplitSUSY}. However, it turns out that the loop hierarchy between gaugino and scalar soft masses is typically reduced or eliminated, as we will see in this Section. 

\begin{figure}
\begin{center}
\begin{picture}(200,100)(50,5)
\PhotonArc(150,50)(40,180,360){3}{8}
\ArrowArc(150,50)(40,0,45)
\ArrowArcn(150,50)(40,90,45)
\ArrowArcn(150,50)(40,180,135)
\ArrowArc(150,50)(40,90,135)
\ArrowLine(70,50)(110,50)
\ArrowLine(230,50)(190,50)
\Text(68,50)[r]{$\lambda^l_a$}
\Text(232,50)[l]{$\lambda^l_b$}
\Text(152,25)[t]{$v^{\mu}$}
\Line(146,94)(154,86)
\Line(154,94)(146,86)
\Text(110,65)[r]{$\lambda^h$}
\Text(192,65)[l]{$\lambda^h$}
\Text(130,90)[r]{$\psi^h$}
\Text(174,90)[l]{$\psi^h$}
\Text(152,98)[b]{$m^{*}_{cd}$}
\Text(151,-10)[b]{$\textrm{(a)}$}
\end{picture}
\begin{picture}(200,100)(50,5)
\DashCArc(150,50)(40,180,360){3}
\ArrowArcn(150,50)(40,90,0)
\ArrowArc(150,50)(40,90,180)
\ArrowLine(70,50)(110,50)
\ArrowLine(230,50)(190,50)
\Text(68,50)[r]{$\lambda^l_a$}
\Text(232,50)[l]{$\lambda^l_a$}
\Text(150,25)[t]{$\phi$}
\Line(146,94)(154,86)
\Line(154,94)(146,86)
\Text(115,75)[r]{$\psi^h$}
\Text(190,75)[l]{$\psi^h$}
\Text(152,98)[b]{$m^{*}_{cd}$}
\Text(151,-10)[b]{$\textrm{(a)}$}
\end{picture}
\\
\begin{picture}(300,100)(0,-10)
\DashCArc(150,50)(40,180,360){3}
\ArrowArcn(150,50)(40,90,0)
\ArrowArc(150,50)(40,90,180)
\ArrowLine(70,50)(110,50)
\ArrowLine(230,50)(190,50)
\Text(68,50)[r]{$\lambda^l_a$}
\Text(232,50)[l]{$\lambda^l_a$}
\Line(146,14)(154,6)
\Line(154,14)(146,6)
\Text(115,75)[r]{$\psi^h$}
\Text(190,75)[l]{$\psi^h$}
\Text(113,25)[r]{$\phi^h$}
\Text(188,25)[l]{$\phi^h$}
\Text(151,28)[t]{$F_{ij}$}
\Text(151,-10)[b]{$\textrm{(b)}$}
\end{picture}
\end{center}
\caption{One loop contributions to light gaugino masses from the exchange of heavy vector (a) and chiral (b) degrees of freedom.}
\label{fig:mg}
\end{figure}
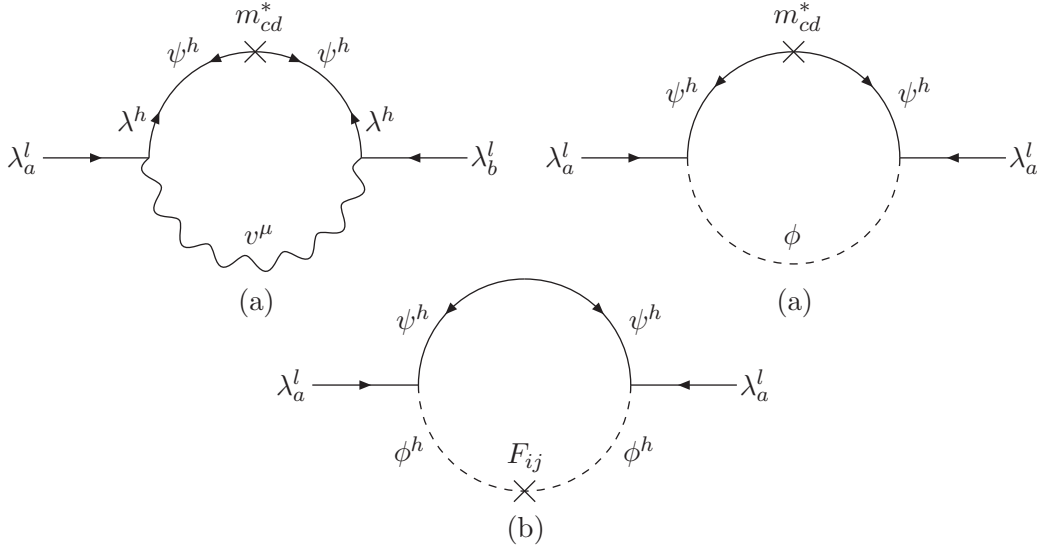

We calculate gaugino masses in the full theory above $M_V$. There are two types of one loop diagrams contributing to gaugino masses, depending on whether the degrees of freedom running in the loop are components of the heavy vector superfields (including the Goldstone superfields), as in Fig.~\ref{fig:mg}a, or physical chiral superfields, as in Fig.~\ref{fig:mg}b. Correspondingly, we will distinguish a ``vector'' and a ``chiral'' contribution to the light gaugino masses,
\begin{equation}
\label{eq:gauginos}
M^g_{ab} = (M^g_{ab})_V + (M^g_{ab})_\Phi .
\end{equation}
The latter may easily dominate on the former, as we will see. The source of supersymmetry breaking entering the diagrams of Fig.~\ref{fig:mg}a and 2b is a tree level splitting among the components of the heavy vector and chiral superfields respectively. We now examine the two contributions in \eq{gauginos} in turn and write the known results~\cite{Hisano:1993zu} in a form general enough to be suitable for the following discussion of their quantitative importance compared to the tree level scalar soft terms. 

\subsection{Vector contribution to gaugino masses}
\label{sec:gauginovector}

In the supersymmetric limit, the fields $v^\mu_a$, $\lambda_a$, $\psi^G_a$, $\re(\phi^G_a)/\sqrt{2}$ form a massive vector multiplet with mass $M_{V_a}$. Once supersymmetry is broken, this spectrum is split by corrections to the fermion and scalar masses, which may also mix them with the components of the physical chiral superfields. Here, we are interested to the supersymmetry breaking fermion mass term in the form $-m_{ab} \psi^G_a \psi^G_b /2$, which is the source of the vector contribution to gaugino masses through the diagrams in Fig.~\ref{fig:mg}a. 

The mass term
\begin{equation}
\label{eq:mab}
m_{ab} = 
\frac{\partial^2 W}{\partial \Phi^G_a \partial \Phi^G_b} (\phi_0)
\end{equation}
vanishes in the supersymmetric limit because of the gauge invariance of $W$. The situation is different in the presence of supersymmetry breaking, when the gauge invariance of $W$ gives 
\begin{equation}
\label{eq:Winvariance}
m_{ab} = g^2 \frac{F^\dagger_0 \{ T^h_a , T^h_b \} \phi_0}{M_{V_a} M_{V_b}} .
\end{equation}
Note also the more general expression for the mixed supersymmetry breaking terms
\begin{equation}
\label{eq:mixing}
\frac{\partial^2 W}{\partial \Phi_i \partial \Phi^G_a} (\phi_0) = \sqrt{2} g 
\frac{F^\dagger_{0j} (T^h_a)_{ji}}{M_{V_a}} .
\end{equation}

Before showing the expression for the gaugino masses induced by $m_{ab}$, let us remind that the heavy vector representation is in general reducible, under the unbroken gauge group $H$, to a set of irreducible components, each with a single value of the mass. Let us call $\hat M_{V_r}$ the value of the mass in the representation $r$
and denote 
\begin{equation}
\label{eq:dMvector}
g^2 \phi^\dagger_0 \{ T^h_a , T^h_b \} F_0 = m^*_{ab} \hat M^2_{V_r} \equiv \frac{\partial \hat M^2_{V_r}}{\partial Z} |F_{0}| \delta_{ab} ,
\end{equation}
if $T^h_a$, $T^h_b$ belong to the representation $r$. 
In the limit $|F_0| \ll M^2_V$, the supersymmetry breaking source $m_{ab}$ can be treated as a perturbation in the one loop computation of gaugino masses. At the leading order in $m_{ab}$, the diagram in Fig.~\ref{fig:mg}a generates a contribution to light gaugino masses given by
\begin{equation}
\label{eq:gauginovector}
(M^g_{ab})_V = -2 
\frac{g^2}{(4\pi)^2} \sum_r S_{ab}(r) \frac{|F_{0}|}{\hat M^2_{V_r}}
\frac{\partial \hat M^2_{V_r}}{\partial Z} ,
\end{equation}
where $S_{ab}(r) = \tr( r(T^l_a) r(T^l_b) )$ is the Dynkin index of the representation $r: T\to r(T)$ of the generator $T$. The above contribution to gaugino masses arises at the scale $M_V$ where the heavy vectors live. 

Let us now discuss the relevance of the above contribution to gaugino masses. First, let us note that in order for $(M^g_{ab})_V$ to be non-vanishing we need the following two conditions to be verified at the same time
\begin{equation}
\label{eq:con1}
\phi^\dagger_0 \{ T^h_{a} , T^h_{b} \} F_0 \neq 0 \text{ (for some $a,b$),} \qquad
\phi^\dagger_0 T^h_{a} F_0 = 0 \text{ (for all $a$)},
\end{equation}
as it can be seen from eqs.~(\ref{eq:dMvector}) and~(\ref{eq:Fgauge}). In particular, we need at least one irreducible (under the full group $G$) chiral superfield multiplet to get vev in both its scalar and $F$ components. At the same time, we need 
\begin{equation}
\label{eq:con2}
F^\dagger_0 T^h_{a} F_0 \neq 0 \text{ (for some $a$)} 
\end{equation}
in order for the tree level contribution to scalar masses to be generated. The conditions in eqs.~(\ref{eq:con1},\ref{eq:con2}) may force the vector contribution to gaugino masses to vanish. On top of that, $(M^g_{ab})_V$ is always suppressed by a loop factor $g^2/(4\pi)^2$ compared to the typical scalar mass in \eq{mm}. If $(M^g_{ab})_V$ was the only contribution to gaugino masses, this would lead to an hierarchy between gaugino and scalar soft masses. Moreover, the present experimental lower limits on gaugino masses, $M_g \gtrsim 100\GeV$, would force the sfermion masses to be heavier than $\ord{10\TeV}$. However, as we will see in a moment, the chiral contribution to gaugino masses can be significantly larger than the vector contribution, thus reducing or even eliminating the loop suppression with respect to soft scalar masses. In this case, the vector contribution to gaugino masses typically ends up to be subdominant. 

\subsection{Chiral contribution to gaugino masses}
\label{sec:gauchi}

The chiral contribution to gaugino masses arises from the one loop diagram in Fig.~\ref{fig:mg}b, as in ordinary loop gauge mediation. The scalar and fermion components of the chiral superfields entering the loop are split by a supersymmetry breaking scalar mass term $-(F_{ij} \phi^h_i \phi^h_j +\text{h.c.})/2$. As a consequence of \eqs{hide} and~(\ref{eq:pure}), the supersymmetry breaking couples directly only to the heavy chiral fields and $F_{ij}$ can be treated as a perturbation in the calculation of gaugino masses. The mass term $F_{ij}$ is then given by
\begin{equation}
\label{eq:Fij}
F_{ij} = -\frac{\partial^3 \hat W}{\partial \Phi^h_i \partial \Phi^h_j \partial Z} (0) |F_{0}| ,
\end{equation}
which adds to the supersymmetric scalar mass term $-M^2_i |\phi'_i|^2$ (in the notation of \eq{MC0diag}). 

The physical chiral superfield representation under the unbroken gauge group $H$ is in general reducible to a set of irreducible components, each with its own mass $\hat M_r$. 
Let us denote 
\begin{equation}
\label{eq:dMchiral}
\frac{\partial^3 \hat W}{\partial \Phi^h_{i} \partial \Phi^h_{j} \partial Z} (0) |F_{0}| =
- F_{ij} \equiv
\frac{\partial \hat M^h_{r}}{\partial Z} |F_{0}| \delta_{ij}.
\end{equation}
At the leading order in $F_{ij}$, the diagram in Fig.~\ref{fig:mg}b generates a contribution to light gaugino masses given by
\begin{equation}
\label{eq:gauginochiral}
(M^g_{ab})_\Phi =  
\frac{g^2}{(4\pi)^2} \sum_r S_{ab}(r) \frac{|F_{0}|}{\hat M_r}
\frac{\partial \hat M_{r}}{\partial Z} .
\end{equation}
Each of the contributions in the sum in the RHS of \eq{gauginochiral} arises at at the scale $\hat M_r$ at which the corresponding chiral superfield lives. If the heavy chiral superfields split in two conjugated representations $\Phi^h = \Psi + \overline\Psi$ with a mass term in the form $-\overline\Psi \mu \Psi$, \eq{gauginochiral} still holds with $M\to\mu$ and a factor 2 multiplying the RHS. 

Let us now discuss the size of the typical chiral contribution to gaugino masses $M_g$ and compare it with the typical size of the tree level scalar soft masses $\tilde m^2$ in \eq{mm}. Let us consider for simplicity the case in which the scalar masses are due to the exchange of a single heavy vector and the irreducible (under $H$) components of the physical chiral superfields have definite charges $Q_r$ under the corresponding generators. As for the dynamics giving rise to gaugino masses, let us assume that there are no bare mass terms in the superpotential, i.e.\ $(\partial^2 W)/ (\partial \Phi_i \partial \Phi_j)(\phi = 0) = 0$. Then both $\hat M_r = \lambda_{rs} \phi_{0s}$ and $(\partial \hat M_{r})/(\partial \phi_{0s}) = \lambda_{rs}$ arise from the same trilinear term in $W(\Phi)$. Under the above assumptions, we have
\begin{equation}
\label{eq:comparison}
\tilde m^2 = \frac{\sum_r (Q_r/Q) |F_{0r}|^2}{\sum_r (Q_r/Q)^2  |\phi_{0r}|^2}
\qquad
M_g = \frac{g^2}{(4\pi)^2} \sum_r S(r) 
\frac{\sum_s \lambda_{rs} F_{0s}}{\sum_s\lambda_{rs} \phi_{0s}} ,
\end{equation}
where $Q$ is the charge of the scalar acquiring the mass $\tilde m$. While the loop factor $g^2/(4\pi)^2$ suppresses $M_g$ compared to $\tilde m$ by a $\ord{100}$ factor, the expressions in \eqs{comparison} may give rise to several enhancements of $\tilde m/M_g$ reducing or even eliminating the loop hierarchy: \begin{itemize}
\item
In the context of grand unified theories the heavy vectors contributing to the soft scalar masses are either one (as in the case of the minimal possibility SO(10), see Section~\ref{sec:guidelines}) or a few, unless the unified group is very large. On the other hand, the gaugino masses may always get contribution from several chiral messengers. 
\item
Sfermion and gaugino masses depend on different group factors. Sfermions can get a mild suppression if $Q_r/Q > 1$. This is indeed what turns out to happen in simple models, as we will see in Section~\ref{sec:guidelines}. 
\item
The heavy vector masses whose exchange generates $\tilde m$ collect all the vevs breaking the corresponding charge $Q$. The scalar mass $\tilde m$ is therefore suppressed by all such vevs. On the other hand, gaugino masses are only suppressed by the vevs related to supersymmetry breaking by superpotential interactions $\lambda_{rs}$. Unless some of then have $Q = 0$, the vevs suppressing gaugino masses will be a subset of the vevs suppressing scalar masses, thus leading to an enhancement of gaugino masses. In the presence of an hierarchy between the vevs related to supersymmetry breaking and some of the other, $Q$-breaking vevs, this enhancement can be quite large. 
\item
Different couplings $\lambda_{rs}$ can appear in the numerator and denominator of the expression $(\sum_s \lambda_{rs} F_{0s})/(\sum_s\lambda_{rs} \phi_{0s})$. This is likely to be the case as a consequence of the relation $\sum_s Q_s (F^*_{0s} \phi_{0s}) = F^\dagger_0 T^h \phi_0 = 0$, which can be satisfied without the need of cancellations only in the case in which the fields charged under $Q$ do not have vevs in both the $F$ and scalar components. If the couplings appearing in the numerator and the denominator are hierarchical, gaugino masses can be sizably enhanced (or, in this case, further suppressed). 
\end{itemize}

The study of simple models shows that indeed the enhancement factors above can naturally arise (see Section~\ref{sec:1610} below and~\cite{NRZ}). In particular, the first two factors reduce the hierarchy $\tilde m/M_g$ by factors 3 and $\sqrt{5}$ respectively. The third factor gives at least a factor $\sqrt{2}$ enhancement. The $\ord{100}$ hierarchy arising from the loop factor is thus reduced by a factor 10, which is enough to bring the sfermions within the reach of the LHC. Such a milder hierarchy may even be necessary in the light of the bounds on the Higgs mass, which require the stops not to be too light. Still, the residual $\ord{10}$ hierarchy can then be easily fully eliminated by the remaining two factors in the above list. 

\subsection{Other one-loop contributions to soft masses}

Besides gaugino masses, which can be seen to arise from one loop corrections to the gauge kinetic function, a number of soft terms can be generated or get a contribution from the one-loop corrections to the K\"ahler. The latter can be computed by using the general results in~\cite{Keff}, which give 
\begin{equation}
\label{eq:dKloop}
\delta_{\text{1-loop}} K = -\frac{1}{32\pi^2}
\left[
\tr\left[
M^\dagger_\Phi M^{\phantom{\dagger}}_\Phi \left(
\log \frac{M^\dagger_\Phi M^{\phantom{\dagger}}_\Phi}{\Lambda^2} - 1
\right)
\right] - 2
\tr\left[
M^2_V \left(
\log \frac{M^2_V}{\Lambda^2} - 1
\right)
\right]
\right] ,
\end{equation}
where 
\begin{equation}
\label{eq:Mfunction}
(M_\Phi)_{ij} = \frac{\partial^2 W}{\partial \Phi_i \partial \Phi_j} (\Phi), \qquad
(M^2_V)_{ab} = \frac{\partial^2 K}{\partial V_a \partial V_b} (\Phi,V=0)
\end{equation}
are functions of the chiral superfields, $K$ is our canonical K\"ahler $K = \Phi^\dagger e^{2gV} \Phi$ and the indexes run on the heavy vector and chiral superfields. As in the case of gaugino masses, the soft terms might get a contribution from both heavy vector and chiral superfields running in the loop. 

As the contribution to one loop soft terms are highly model dependent, we just remind and collect their general expression in terms of $\delta_{\text{1-loop}} K$. Let us expand $\delta_{\text{1-loop}} K$ in terms of powers of $Q$ and $Z$ around $\phi_0$. The relevant terms are
\begin{equation}
\label{eq:dKloop2}
\delta_{\text{1-loop}} K = 
\left(\alpha^{(1)}_{ij} Z Q_i^{\dagger} Q_j + \frac{\beta^{(1)}_{ij}}{2} Z^{\dagger} Q_i Q_j + \text{h.c.}
\right) + 
\alpha^{(2)}_{ij} Z^{\dagger} Z {Q_i}^{\dagger} Q_j + 
\left(\frac{\beta^{(2)}_{ij}}{2} Z^{\dagger} Z {Q_i} Q_j + \text{h.c.} \right)
+\ldots ,
\end{equation}
where $\alpha^{(1)}$, $\alpha^{(2)}$, are hermitian, $\beta^{(1)}$, $\beta ^{(2)}$ symmetric and all are dimensionful. We have omitted $Z^{\dagger}Q_i$ terms, which are well-known to destabilize the hierarchy~\cite{PolchinskiSusskind}. Their absence can be ensured for example by requiring that there are no light chiral fields with the same quantum numbers as $Z$.

The first term $\alpha^{(1)}$ gives rise to the following ``$A$-terms''
\begin{equation}
\label{eq:A}
\mathcal{L}^A_\text{1-loop} = -A_{ij} q_i 
\frac{\partial \hat W}{\partial Q_j} (q), \quad\text{with}\quad
A_{ij} = |F_0| \alpha^{(1)} 
\end{equation}
(and to a two loop contribution to scalar soft masses), where $q$ is the scalar component of $Q$. The second term $\beta^{(1)}$ generates a contribution to the ``$\mu$-term'' in the superpotential
\begin{equation}
\label{eq:mu}
W^\mu_\text{1-loop} = \frac{\mu_{ij}}{2} Q_i Q_j, \quad\text{with}\quad
\mu_{ij} = |F_0| \beta^{(1)} ,
\end{equation}
and the fourth term  $\beta ^{(2)}$ a contribution to the ``$B_\mu$-term''  
\begin{equation}
\label{eq:Bmu}
\mathcal{L}^{B_\mu}_\text{1-loop} = -\frac{(B_\mu)_{ij}}{2} q_i q_j , \quad\text{with}\quad
(B_\mu)_{ij} = -|F_0|^2 \beta^{(2)} .
\end{equation}
A more comprehensive discussion of the $\mu$-term and the $\mu$ problem can be found in Section~\ref{sec:mu}.

Finally, $\alpha^{(2)}$ gives 1-loop contributions to soft scalar masses 
\begin{equation}
\label{eq:dmm}
\delta \tilde m^2_{ij} = -|F_0|^2 \alpha^{(2)}_{ij}
\end{equation}
that add to the tree level contributions in \eq{mm}. 

Additional one-loop contributions to soft scalar masses can come from an induced 1-loop Fayet-Iliopoulos term~\cite{Multimess} associated for example to the heavy $H$-singlet generators, in particular to those involved in the mediation of supersymmetry breaking at the tree level. Such terms vanish if the heavy chiral mass matrix and the matrix of their couplings to the spurion $Z$ are diagonal in the same basis (in which case the condition in \eq{pure} is also automatically satisfied) or if the latter matrix of couplings is hermitian in one basis in which the mass matrix is diagonal~\cite{Dvali:1996cu}. 

This completes the list of the soft terms arising at one loop. Two loop corrections to the scalar soft masses can also arise, of course, as in standard loop gauge mediation, and be sizable in the presence of an enhancement of one-loop gaugino masses~\cite{NRZ}. 

\section{Guidelines for model building}
\label{sec:guidelines}

We now consider the possibility to obtain a phenomenologically viable model from the general formalism discussed so far. We will see that clear model building guidelines emerge from this analysis, leading, in economical schemes, to peculiar predictions for the pattern of MSSM sfermion masses. In particular, we will identify the assumptions underlying such predictions. 

In a phenomenologically viable model, the unbroken gauge group $H$ should contain the SM group, $\GSM\subseteq H$, and the light superfield content should contain the 
MSSM spectrum, $(q_i,u^c_i,d^c_i,l_i,e^c_i) \subseteq Q$, in standard notations, where $i=1,2,3$ is the family index. We assume that the full gauge group $G$ is a simple, grand-unified group, motivated by the well known successful predictions of the SM fermion gauge quantum numbers, of the strong coupling in the MSSM, and of the unification scale in the phenomenologically allowed region.
The candidates for the unified group $G$ in a four-dimensional theory are SU($N$), $N\geq 5$, SO(4$n$+2), $n\geq 2$, and the exceptional group $E_6$~\cite{Slansky:1981yr}. In the following we will focus on the smallest (or unique) representatives of each class, SU(5), SO(10), and $E_6$. 

We want the MSSM sfermions to get a positive, $\ord{\TeV}$ mass through tree level gauge mediation. The general form of such mass terms is given in \eq{mm}. The latter contains two contributions, corresponding to the two diagrams on the right-hande side in \Fig{diagram}. In order for the second contribution to play a role for sfermion masses, the corresponding chiral superfields should live in the same unified multiplet as the supersymmetry breaking source $Z$. This will not be the case in the models we consider (as a consequence, for example, of a matter parity telling the supersymmetry breaking multiplet from the matter ones). On the other hand, the second contribution might contribute to the Higgs masses, if some of the gauge generators have the same quantum numbers (which is not the case in SO(10), the unified group we will consider in greater detail). 

The MSSM sfermions then get their tree level soft masses from the first term in \eq{mm} only. In order for $F_0^\dagger T^h_a F^{\phantom{\dagger}}_0$ to be non-vanishing, the heavy generator $T^h_a$ must be a SM singlet, since $F_0$ is. We therefore need a group $G$ with rank 5 at least. This means that SU(5) cannot give rise to tree level gauge mediation, while SO(10) and $E_6$ are in principle suitable. 

Let us first consider the ``minimal'' option, SO(10), which has also the well known virtue to be able to accommodate a whole MSSM family in a single irreducible spinorial representation. We will make a few considerations on the $E_6$ option at the end of this Section. In SO(10) there is exactly one (up to a sign) ortho-normalized heavy SM-singlet generator, $T_h = 1/\sqrt{40} X$, where $X = 5(B-L) -4Y$ is the SU(5) invariant SO(10) generator. The quantum numbers of the SO(10) representations with dimension $d<120$ under $X$ are given in Table~\ref{tab:X}. 
\begin{table}
\begin{equation*}
\begin{array}{|c|ccc|ccc|cc|cccc|ccc|}
\hline
\text{SO(10)} &  \multicolumn{3}{c|}{16} & \multicolumn{3}{c|}{\overline{16}} & \multicolumn{2}{c|}{10} & \multicolumn{4}{c|}{45} & \multicolumn{3}{c|}{54} \\ \hline
\text{SU(5)} & 1 & 10 & \overline{5} & 1 & \overline{10} & 5 & 5 & \overline{5} & 1 & 10 & \overline{10} & 24 & 24 & 15 & \overline{15} \\ \hline
X & 5 & 1 & -3 & -5 & -1 & 3 & -2 & 2 & 0 & -4 & 4 & 0 & 0 & -4 & 4 \\
\hline
\end{array}
\end{equation*}
\caption{Quantum numbers of the non-trivial SO(10) representations with dimension $d< 120$ under the SO(10) generator $X$.}
\label{tab:X}
\end{table}
The values of the $X$ quantum numbers are crucial because the soft terms turn out to be proportional to those charges. From \eq{mm} we obtain in fact
\begin{equation}
\label{eq:mmf}
\tilde m^2_f = \frac{X_f (F_0^\dagger X F^{\phantom{\dagger}}_0)}{\phi_0^\dagger X^2 \phi^{\phantom{\dagger}}_0} \quad
\text{at the scale} \quad
M_V = \frac{g^2}{20} \phi_0^\dagger X^2 \phi^{\phantom{\dagger}}_0,
\end{equation}
where $X_f$ is the $X$-charge of the sfermion $\tilde f$ and $M_V$ is the mass if the vector superfield associated to the generator $X$ (note that the gauge coupling and the normalization of the generator $T_h$ cancel in \eq{mmf}). In order to predict the pattern of the tree level sfermion masses, we then just need to specify the embedding of the three MSSM families into SO(10), which we will do through their SU(5) embedding into three light $\bar 5^l_i + 10^l_i$, $i=1,2,3$. 

We use two constraints to determine the embedding of the $\overline{5}^l_i + 10^l_i$ into SO(10) representations. The first one is related to quite a nice feature \eq{mmf}: the soft terms turn out to be family-universal, thus neatly solving the supersymmetric flavor problem. Provided, of course, that the three families of each of the MSSM matter multiplets are embedded in the same type of SO(10) representation, which we will assume in order to ensure that family-universality indeed holds. On top of that, we want the MSSM sfermion soft masses in \eq{mmf} to be positive in order to avoid spontaneous symmetry breaking of color, electric charge, or lepton number at the scale $\tilde m$. Clearly, the standard embedding of a whole family into a 16 of SO(10) would not work, as it would lead to negative masses for the sfermions in either the $\bar 5$ or the 10 of SU(5). This is in turn related to the 
tracelessness of the SO(10) generators, and in particular of $X$. As a consequence, whatever is the SO(10) representation in which we choose to embed a given MSSM matter multiplet with positive soft mass, that representation will necessarily contain extra fields with negative soft masses. This apparent obstacle can be easily overcome by splitting the SO(10) representation containing the MSSM multiplet through SO(10) breaking, in such a way that the extra fields with negative soft masses acquire a large supersymmetric mass term. The negative soft mass will then represent a negligible supersymmetry breaking correction to that large (positive) mass. It turns out that such a splitting is actually expected to arise, as will see, a fact that reinforces the logical consistency of tree level gauge mediation. 

We are now ready to discuss the embeddings of the three $\overline{5}^l_i$ and $10^l_i$ of SU(5) containing the light MSSM families in SO(10). As $\phi_0^\dagger X^2 \phi^{\phantom{\dagger}}_0$ is positive, the possible choices depend on the sign of $F_0^\dagger X F^{\phantom{\dagger}}_0$. We limit ourselves to the SO(10) representations with $d<120$, as in Table~\ref{tab:X}. There are then only two possibilities:
\begin{itemize}
\item
$F_0^\dagger X F^{\phantom{\dagger}}_0>0$. In this case we need to embed the $\overline{5}^l_i$'s and $10^l_i$'s into SO(10) representations containing $\overline{5}$ and $10$ of SU(5) with positive charges under $X$. From Table~\ref{tab:X} we see that the only possibility is to use three $16_i = (1^{16}_i, 10^{16}_i,\overline{5}^{16}_i)$ and three $10_i = (5^{10}_i,\overline{5}^{10}_i)$, $i=1,2,3$, where we have explicitly indicated the SU(5) decomposition, and to embed the  $10^l_i$'s into the $16_i$'s, $10^l_i \equiv 10^{16}_i$, and the $\overline{5}^l_i$'s into the $10_i$'s, $\overline{5}^l_i \equiv \overline{5}^{10}_i$. The spare components $\overline{5}^{16}_i$, $5^{10}_i$ get negative soft masses and need to acquire a large supersymmetric mass term. 
\item
$F_0^\dagger X F^{\phantom{\dagger}}_0<0$. In this case we need the $\overline{5}^l_i$'s and $10^l_i$'s to have negative charges under $X$. The only possibility is then to use three $16_i$'s as before and  three $45_i = (1^{45}_i, 10^{45}_i,\overline{10}^{45}_i,24^{45}_i)$, $i=1,2,3$, with $\overline{5}^l_i \equiv \overline{5}^{16}_i$ and $10^l_i \equiv 10^{45}_i$. The spare components $10^{16}_i$, $\overline{10}^{45}_i$, get negative or vanishing soft masses and need to acquire a large supersymmetric mass term. 
\end{itemize}
In both cases the chiral content of the theory is still given by three 16 of SO(10). We have implicitly neglected the possibility of mixed embeddings in which, for example, the $\bar 5_i$'s of SU(5) are a superposition of the $\bar 5_i$'s in the $10_i$'s and $16_i$'s of SO(10). While this possibility is in principle not excluded, it would in general introduce a dependence of the sfermion soft masses on mixing parameters that are in general flavor violating, thus possibly spoiling the flavor universality result. 

The two possibilities above give rise to two definite predictions for the patter of sfermion soft masses at the scale $M_V$:  
\begin{gather}
\label{eq:prediction}
(\tilde m^2_{l})_{ij} = (\tilde m^2_{d^c})_{ij} = m^2_{\overline{5}} \delta_{ij}, \quad
(\tilde m^2_{q})_{ij} = (\tilde m^2_{u^c})_{ij} = (\tilde m^2_{d^c})_{ij} = m^2_{10} \delta_{ij}, \quad\text{with} \notag \\[2mm]
m^2_{\overline{5}} = 2 m^2_{10} \quad\text{if} \quad F_0^\dagger X F^{\phantom{\dagger}}_0>0 \\
m^2_{\overline{5}} = \frac{3}{4} m^2_{10} \quad\text{if} \quad F_0^\dagger X F^{\phantom{\dagger}}_0<0 . \notag
\end{gather}
To summarize, the latter predictions are based on the following hypotheses: ``minimal'' unified gauge group SO(10), embedding of the MSSM families in the SO(10) representations with dimension $d<120$ not containing the Goldstino, and absence of mixed embeddings to automatically preserve flavor-universality.  The predictions on the ratios $m_{\overline{5}}/m_{10}$ in \eq{prediction} are peculiar enough to make a possible experimental test at the LHC a strong hint for tree level gauge mediation. 

As for the source of supersymmetry breaking, $\vev{Z} = |F_0| \theta^2$, we need $Z$ to have a non-vanishing charge under $X$. If we limit ourselves again to representations with $d<120$, the only possibility is that $Z$ has a component in the ``right-handed neutrino'' direction of a 16 or a $\overline{16}$. With the sign conventions we adopted, a component in a 16 gives a positive contribution to $F_0^\dagger X F^{\phantom{\dagger}}_0$, while a component in a $\overline{16}$ gives a negative contribution. 

\interskip

We now want to show that the two embeddings of the light MSSM families described above can be obtained in a natural way. We have to show that it is possible to split the SO(10) representations in which the MSSM fields are embedded in such a way that the extra fields (with negative soft masses) get a heavy supersymmetric mass term from SO(10) breaking. It will turn out that the SO(10) breaking vevs of a $16+\overline{16}$, essential to break SO(10) to the SM (unless representations with $d \geq 126$ are used to reduce the rank) just provide the needed splitting. The fact that such vevs make heavy precisely the components of the SO(10) representations that get a negative soft supersymmetry breaking mass reinforces the logical consistency of this framework.  In the following, we first discuss the $16_i+10_i$ embedding in a general, top-bottom perspective, obtaining a generalization of the model in~\cite{NRZ}, and discuss the conditions for a pure (non mixed) embedding. We then discuss the possibility of a $16_i+45_i$ embedding. 

\subsection{The embedding into $16_i+10_i$, $i=1,2,3$}
\label{sec:1610}

Let us consider the embedding associated to the case $F_0^\dagger X F^{\phantom{\dagger}}_0 > 0$. We assume the existence of a matter parity symmetry that tells matter superfields from Higgs superfields. Let 16, $\overline{16}$ be the SO(10) multiplets breaking SO(10) to SU(5) (we can always choose the basis in the space of the 16 ($\overline{16}$) representations in which a single 16 ($\overline{16}$) gets a vev in its scalar component). The most general renormalizable superpotential involving 16, $\overline{16}$, $16_i$, $10_i$, $i=1,2,3$, and invariant under a matter parity under which the SO(10) Higgs fields 16, $\overline{16}$ are even and the the matter fields are odd is
\begin{equation}
\label{eq:W1}
W = h_{ij} 16_i 10_j 16 + \frac{\mu_{ij}}{2} 10_i 10_j + W_\text{vev} ,
\end{equation}
where $W_\text{vev}$ takes care of providing a vev to the 16, $\overline{16}$ in the SM-singlet direction and does not depend on the matter fields (but can involve additional even fields\footnote{The simplest possibility is $W_\text{vev} = X(\overline{16} 16 -V^2)$, where $X$ is an SO(10) singlet.}). The term $h_{ij} 16_i 10_j 16$ is just what needed to split the SU(5) components of the $16_i = (1^{16}_i, 10^{16}_i,\overline{5}^{16}_i)$ and of the $10_i = (5^{10}_i,\overline{5}^{10}_i)$ and make heavy the unwanted components $\overline{5}^{16}_i$ and $5^{10}_j$. Once 16 acquires a vev $V$ in its singlet neutrino component, in fact, a mass term is generated for those components,
\begin{equation}
\label{eq:unwanted1}
M_{ij} \overline{5}^{16}_i 5^{10}_j ,\qquad M_{ij} = h_{ij} V .
\end{equation}
The singlet neutrinos $1^{16}_i$ remain light at the renormalizable level but can get a mass at the non-renormalizable level through the operator $(\overline{16} 16_i)(\overline{16} 16_j)/\Lambda$. 

It is remarkable that the components acquiring a large mass are precisely those that get a negative soft mass term. On the other hand, this is only true in the limit in which the $\mu_{ij}$ mass term in \eq{W1} can be neglected. In the presence of a non negligible $\mu_{ij}$, in fact, the full mass term would be
\begin{equation}
\label{eq:mixing1}
(\overline{5}^{16}_i M_{ij} + \overline{5}^{10}_i \mu_{ij} ) 5^{10}_j ,
\end{equation}
which would give rise to a mixed embedding of the light $\overline{5}^l_i$'s in the $16_i$'s and $10_i$'s. In order to abide to our assumptions, which exclude the possibility of mixed embeddings, such a $\mu_{ij}$ term should be absent. This can be easily forced by means of an appropriate symmetry. Let us however relax for a moment that assumption in order to quantify the deviation from universality associated to a small, but non-negligible $\mu_{ij}$. The MSSM sfermions in the $\overline{5}$ of SU(5) receive in this case two contributions to their soft mass, a positive one associated to the components in the $10_i$'s, proportional to $X(\overline{5}^{10}) = 2$, and a negative one associated to the components in the $16_i$'s, proportional to $X(\overline{5}^{16}) = -3$. The soft mass matrix for the light sfermions in the $\overline{5}$ of SU(5) can be easily calculated in the limit in which the $\mu_{ij}$ mass term can be treated as a perturbation. In this limit, the light MSSM fields in the $\overline{5}$ of SU(5) are in fact  
\begin{equation}
\label{eq:sup1}
\overline{5}^l_i \approx \overline{5}^{10}_i - (\mu M^{-1})^*_{ij} \overline{5}^{16}_j 
\end{equation}
and their soft scalar mass matrix at the scale $M_V$ is 
\begin{equation}
\label{eq:mmsuperpos1}
(\tilde m^2_{\overline{5}})_{ij} \approx \frac{2}{5}
\tilde m^2 
\left(
\delta_{ij} - \frac{5}{2} \left(
\mu^* M^{*-1} M^{T-1} \mu^T
\right)_{ij}
\right) ,
\end{equation}
where $\tilde m^2$ is defined below. The mixed embedding induced by the mass term $\mu_{ij}$ leads to flavor-violating soft-terms. Setting $\mu_{ij}=0$ allows to preserve the flavor blindness of the soft terms and to satisfy  the FCNC constraints without the need of assumptions on the structure of the flavor matrices $h_{ij}$ and $\mu_{ij}$. We therefore assume that $\mu_{ij}$ is vanishing or negligible. We then have $\overline{5}^l_i = \overline{5}^{10}_i$, $10^l_i = 10^{16}_i$, with the extra components $\overline{5}^{16}_i$ and $5^{10}_i$ obtaining a large supersymmetric mass term $M_{ij} \overline{5}^{16}_i 5^{10}_i$, as desired. The soft masses for the light sfermions are \globallabel{eq:prediction1}
\begin{gather}
(\tilde m^2_{l})_{ij} = (\tilde m^2_{d^c})_{ij} = \frac{2}{5} \tilde m^2 \delta_{ij}, \quad
(\tilde m^2_{q})_{ij} = (\tilde m^2_{u^c})_{ij} = (\tilde m^2_{d^c})_{ij} = \frac{1}{5} \tilde m^2 \delta_{ij}  ,
\mytag \\ \text{with}\quad 
\tilde m^2 = 5\frac{(F_0^\dagger X F^{\phantom{\dagger}}_0)}{\phi_0^\dagger X^2 \phi^{\phantom{\dagger}}_0} >0 , \mytag 
\end{gather}
as anticipated in \eq{prediction}. The reason for the factor $5=X(1^{16})$ will become clear in a moment. 

We now need to identify the embedding of the MSSM Higgs superfields and obtain the MSSM superpotential for them, in particular the MSSM Yukawa interactions. It is useful to discuss the Yukawa interaction in SU(5) language. The up quark Yukawa interactions arise from the SU(5) operator
\begin{equation}
\label{eq:l1}
\frac{\lambda^{(1)}_{ij}}{2} 10^l_i 10^l_j 5_H ,
\end{equation}
where $5_H$ contains the MSSM up Higgs. As $10^l_i = 10^{16}_i$, the operator in \eq{l1} can arise at the renormalizable level from a SO(10) invariant operator only if $5_H$ has a component into a $10_H$ of SO(10), $10_H = (5^{10}_H,\overline{5}^{10}_H)$, with 
\begin{equation}
\label{eq:cd}
5_H^{10} = \cos\theta_u 5_H + \ldots, \quad 0\leq \theta_u \leq \pi/2,
\end{equation}
where $\cos^2\theta_u$ measures the size of the $5_H$ component from 10 representations of SO(10) (a basis in the space of the 10 representations can always be chosen such that $5_H$ is contained in a single one, the $10_H$). The operator in \eq{l1} will then originate as 
\begin{equation}
\label{eq:y}
\frac{y^H_{ij}}{2} 16_i 16_j 10_H = \frac{\lambda^{(1)}_{ij}}{2} 10^l_i 10^l_j 5_H +\ldots ,
\quad \text{with}\quad
\lambda^{(1)}_{ij} = \cos\theta_u y^H_{ij} .
\end{equation}

The down quark and charged lepton Yukawa interactions arise at the renormalizable level\footnote{SU(5)-invariant renormalizable Yukawa interactions lead to wrong mass relations for the two lighter families of down quarks and charged leptons. This may indicate that the light family Yukawas arise at the non-renormalizable level, as also suggested by their smallness. We ignore this issue in the following and only consider the renormalizable part of the superpotential.} from the SU(5) operator 
\begin{equation}
\label{eq:l2}
\lambda^{(2)}_{ij} 10^l_i \overline{5}^l_j \overline{5}_H ,
\end{equation}
where $\overline{5}_H$ contains the MSSM down Higgs. As $10^l_i = 10^{16}_i$ and $\overline{5}^l_i = \overline{5}^{10}_i$, the operator in \eq{l2} can arise at the renormalizable level from a SO(10) invariant operator only if $\overline{5}_H$ has a component into a $16_H$ of SO(10), $16_H = (1^{16}_H,10^{16}_H,\overline{5}^{16}_H)$, with 
\begin{equation}
\label{eq:cu}
\overline{5}_H^{16} = \sin\theta_d 5_H + \ldots, \quad 0\leq \theta_d \leq \pi/2,
\end{equation}
where $\sin^2\theta_d$ measures the size of the $\overline{5}_H$ component from 16 representations of SO(10). The operator in \eq{l2} will then originate as 
\begin{equation}
\label{eq:h}
h^H_{ij} 16_i 10_j 16_H = \lambda^{(2)}_{ij} 10^l_i \overline{5}^l_j \overline{5}_H +\ldots ,
\quad \text{with}\quad
\lambda^{(2)}_{ij} = \sin\theta_d h^H_{ij} .
\end{equation}
It is tempting (and economical) to identify the $16_H$ with 16, the field whose vev breaks SO(10) to SU(5), in which case $h^H = h$ and the mass of the heavy extra components $\overline{5}^{16}_i$ and $5^{10}_i$ in \eq{unwanted1} turns out to be proportional to the corresponding light fermion masses (up to non-renormalizable corrections needed to fix the light fermion mass ratios)\footnote{This property can give rise to a predictive model of leptogenesis in the context of type-II see-saw models~\cite{FHLR,CFLR}.}. 

Having introduced the MSSM Higgs fields, let us now discuss their soft mass terms. To summarize the previous discussion, with our $d<120$ representation content, the up (down) Higgs superfield $h_u$ ($h_d$) can be embedded in either 10's or 16's ($\overline{16}$'s) of SO(10), in both cases through the embedding into a $5_H$ ($\overline{5}_H$) of SU(5). We have denoted by $\cos^2\theta_u$ ($\cos^2\theta_d$) the overall size of the $h_u$ ($h_d$) component in the 10's. The overall size of the component in the 16's ($\overline{16}$'s) is then measured by $\sin^2\theta_u$ ($\sin^2\theta_d$). Correspondingly, the Higgs soft masses get two contributions from the first term in \eq{mmf} proportional to two different $X$ charges:
\globallabel{eq:mmH}
\begin{gather}
m^2_{h_u} = \frac{-2c^2_u + 3 s^2_u}{5} \tilde m^2, \quad
m^2_{h_d} = \frac{2c^2_d - 3 s^2_d}{5} \tilde m^2, \quad \text{so that} \mytag \\
-\frac{2}{5} \tilde m^2 \leq m^2_{h_u} \leq \frac{3}{5} \tilde m^2, \quad
-\frac{3}{5} \tilde m^2 \leq m^2_{h_d} \leq \frac{2}{5} \tilde m^2 . \mytag 
\end{gather}

Let us now consider gaugino masses. A general discussion of all possible contributions to gaugino masses in the embedding we are considering and in the presence of an arbitrary number of SO(10) representation with $d<120$ would be too involved. We then consider a few examples meant to generalize the case considered in~\cite{NRZ} and  to illustrate the general properties discussed in Section~\ref{sec:oneloop}. 

Let us begin by illustrating in more detail the structure of supersymmetry breaking. With the representation content of Table~\ref{tab:X}, supersymmetry breaking can be associated to the $F$-term vevs of superfields in 16, $\overline{16}$, 45, 54 representations (the ones containing SM singlets). However, only the 16, $\overline{16}$, whose singlets have non-vanishing $X$-charges, can contribute to tree level soft masses. Let us call $16^H_\alpha$, $\overline{16}^H_\alpha$ the matter parity even superfields in the 16 and $\overline{16}$ representations of SO(10). In a generic basis, we can parametrize the vevs of their singlet components as
\begin{equation}
\label{eq:F16}
\vev{1^{16_H}_\alpha} = V_\alpha + F_\alpha \theta^2 \qquad
\big\langle1^{\overline{16}_H}_\alpha\big\rangle = \overline{V}_\alpha + \overline{F}_\alpha \theta^2 .
\end{equation}
The $D$-term condition for the $X$ generator requires
\begin{equation}
\label{eq:XDterm}
\sum_\alpha |V_\alpha|^2 \approx \sum_\alpha |\overline{V}_\alpha|^2 ,
\end{equation}
while gauge invariance gives
\begin{equation}
\label{eq:gaugeinv}
\sum_\alpha V^*_\alpha F_\alpha = \sum_\alpha \overline{V}^{*}_\alpha \overline{F}_\alpha .
\end{equation}
Sfermion masses are proportional to 
\begin{equation}
\label{eq:mmbar}
\tilde m^2 = 
\frac{\sum_\alpha (|F_\alpha|^2 - |\overline{F}_\alpha|^2)}{\sum_\alpha (|V_\alpha|^2 + |\overline{V}_\alpha|^2)}
\end{equation}
(due to the factor 5 in the definition of $\tilde m^2$), where $\sum_\alpha |F_\alpha|^2 > \sum_\alpha |\overline{F}_\alpha|^2$ by definition in the case we are considering. Note that $\tilde m^2$ is suppressed by \emph{all} vevs contributing to $X$ breaking. 

Let us now comment on the vector contribution to gaugino masses. Let us assume to begin with that the $\overline{16}$'s do not break supersymmetry. Without loss of generality we can then assume that supersymmetry breaking is only associated to $16' \equiv 16^H_1$. The gauge invariance condition then gives $V_1 = 0$, i.e.\ a vev for both the $F$-term and scalar components is not allowed. Since the $F$-term and scalar components belong to different irreducible representations, no vector contribution to gaugino masses is generated by the 16's. A vector contribution can still be generated by the $F$-term vev of a 45, for example, for which the gauge invariance condition does now prevent a vev in both the scalar and $F$-term component. Or, it can be generated by the $F$-terms of the 16's if some of the $\overline{16}$ also breaks supersymmetry and cancels the contribution of the 16 to \eq{gaugeinv}.  

Let us next consider the chiral contribution to gaugino masses. The massive components $\overline{5}^{16}_i$ and $5^{10}_j$ of the matter superfields will act as chiral messengers if they are coupled to supersymmetry breaking. Let us then consider as before the case in which the $\overline{16}$'s do not break supersymmetry, supersymmetry breaking is provided by the $F$-term vev $F$ of  the singlet component of the $16'$ and is felt by the chiral messengers through the $h'_{ij} 16_i 10_j 16'$ interaction. Let $16\equiv 16^H_2$ be the field whose vev gives mass to the $\overline{5}^{16}_i$, $5^{10}_j$ through the $h_{ij} 16_i 10_j 16$ interaction, as in \eq{W1}. And let us assume that additional $16^H_\alpha$'s and $\overline{16}^H_\alpha$'s get vevs in their scalar components. The chiral messengers $\overline{5}^{16}_i$, $5^{10}_j$ have therefore a supersymmetric mass $M_{ij} = h_{ij} V$ and their scalar components get a supersymmetry breaking term mass term $F_{ij} = h'_{ij} F$. The induced one loop chiral contribution to gaugino masses is then 
\begin{equation}
\label{eq:gauginoexample}
M_g = \frac{g^2}{(4\pi)^2} \tr(h' h^{-1}) \frac{F}{V} .
\end{equation} 
The tree level soft mass of the stop (belonging to the 10 of SU(5)) is 
\begin{equation}
\label{eq:mstop}
\tilde m^2_t = \frac{1}{5} \frac{|F|^2}{|V|^2 + \sum_\alpha |V_\alpha|^2+ |\overline{V}|^2+ \sum_\alpha |\overline{V}_\alpha|^2} .
\end{equation} 
We can then compare stop and gaugino masses (before radiative corrections). Their ratio is particularly interesting, as the gaugino mass $M_g$ is at present bounded to be heavier than about $100\GeV$, while $\tilde m_t$ enters the radiative corrections to the Higgs mass. Therefore, the ratio $\tilde m_t / M_g$ should not be too large in order not to increase the fine-tuning and not to push the stops and the other sfermions out of the LHC reach. From the previous equations we find 
\begin{equation}
\label{eq:ratio}
\frac{M_g}{\tilde m_t}  = \frac{3\sqrt{5 k}}{(4\pi)^2}\, \lambda , \quad \lambda = \frac{g^2 \tr(h'h^{-1})}{3} ,
\quad k = \frac{|V|^2+ \sum_\alpha |V_\alpha|^2 + |\overline{V}|^2 + \sum_\alpha |\overline{V}_\alpha|^2}{|V|^2} \geq 2.
\end{equation}
\Eq{ratio} illustrates all the enhancement factors discussed in Section~\ref{sec:oneloop} that can compensate the loop suppression of gaugino masses. The factor 3 corresponds to the number of chiral messenger families ($\tr(h'h^{-1}) = 3$ for $h = h'$) contributing to gaugino masses, to be compared to the single vector messenger generating sfermion masses at the tree level. The factor $\sqrt{5}$ comes from the ratio of charges $X(1^{16})/X(10^{16})= 5$ suppressing the stop mass in \eq{comparison}. The factor $k \geq 2$ is the ratio of the vev suppressing gaugino masses (the one related to supersymmetry breaking though superpotential interactions, $|V|^2$), and the combination of vevs suppressing sfermion masses (all of them). Note that in the presence of hierarchies of vevs, the factor $k$ can be large. Finally $\lambda$ represents a combination of couplings that can further enhance (or suppress, in this case) gaugino masses. All in all, we see that the loop factor separating $\tilde m_t$ and $M_g$ is partially compensated by a combination of numerical factors: $(4\pi)^2 \sim 100$ (leading to $\tilde m_t \gtrsim 10\TeV$ for $\lambda=1$) becomes at least $(4\pi)^2/(3\sqrt{10}) \sim 10$ (leading to $\tilde m_t \gtrsim 1\TeV$ for $\lambda=1$). A largish value of the factors $k$ or $\lambda$ can then further reduce the hierarchy and even make $M_g \sim \tilde m_t$, if needed. 

\subsection{The embedding into $16_i+45_i$, $i=1,2,3$}

Let us now consider the second type of embedding identified above, corresponding to $F_0^\dagger X F^{\phantom{\dagger}}_0 < 0$. The most general renormalizable superpotential involving 16, $\overline{16}$ and $16_i$, $45_i$, $i=1,2,3$ and invariant under matter parity is
\begin{equation}
\label{eq:W2}
W = h_{ij} 16_i 45_j \overline{16} + \frac{\mu_{ij}}{2} 45_i 45_j + W_\text{vev} .
\end{equation}

The term $h_{ij} 16_i 45_j \overline{16}$ is just what needed to split the SU(5) components of the $16_i = (1^{16}_i, 10^{16}_i,\overline{5}^{16}_i)$ and of the $45_i = (1^{45}_i, 10^{45}_i,\overline{10}^{45}_i,24^{45}_i)$ and make heavy the unwanted components $10^{16}_i$ and $\overline{10}^{45}_j$. Once 16 acquires a vev $V$, in fact, a mass term is generated for those components,
\begin{equation}
\label{eq:unwanted2}
M_{ij} 10^{16}_i  \overline{10}^{45}_j ,\qquad M_{ij} = h_{ij} V .
\end{equation}

It is remarkable that also in this case the components acquiring a large mass are precisely those that get a negative soft mass term. On the other hand, this is only true in the limit in which the $\mu_{ij}$ mass term in \eq{W2} can be neglected. In order to abide to our pure embedding assumption, we will neglect such a term. Let us note, however, that such a term should arise at some level in order to make the $24^{45}_i$'s components heavy. Note that the $24_i$'s do not affect gauge coupling unification at one loop and can therefore be considerably lighter than the GUT scale, consistently with the required smallness of $\mu_{ij}$. The soft masses for the light sfermions are now
\begin{gather}
\label{eq:prediction2}
(\tilde m^2_{l})_{ij} = (\tilde m^2_{d^c})_{ij} = \frac{3}{5} \tilde m^2 \delta_{ij}, \quad
(\tilde m^2_{q})_{ij} = (\tilde m^2_{u^c})_{ij} = (\tilde m^2_{d^c})_{ij} = \frac{4}{5} \tilde m^2 \delta_{ij}  ,
\\ \text{with}\quad 
\tilde m^2 = -5\frac{(F_0^\dagger X F^{\phantom{\dagger}}_0)}{\phi_0^\dagger X^2 \phi^{\phantom{\dagger}}_0} > 0.
\end{gather}

Unfortunately, the embedding we are discussing cannot be implemented with renormalizable interactions and $d < 120$ representations only. The problem is obtaining the Yukawa interactions. Let us consider the up quark Yukawas, arising as we saw from the SU(5) operator in \eq{l1}. Given its size, we expect at least the top Yukawa coupling to arise at the renormalizable level. As in the present case $10^l_i = 10^{45}_i$, the operator in \eq{l1} can arise at the renormalizable level from a SO(10) invariant operator only if $5_H$ has a component in a SO(10) representation coupling to $45_i 45_j$. And the lowest dimensional possibility containing the 5 of SU(5) is the 210. For this reason, we do not pursue this possibility further here, although models with large representations are not a priori excluded. 

\subsection{$E_6$}

We close this Section with a few considerations about the possibility to identify the unified group with $E_6$. Such a possibility looks particularly appealing in the light of what above. We have seen in fact that the most straightforward possibility to realize tree level gauge mediation in SO(10) requires the matter superfield content to include three $16_i + 10_i$, $i=1,2,3$. This is precisely what $E_6$ predicts. The fundamental of $E_6$, in fact, a representation of dimension 27, decomposes as 
\begin{equation}
\label{eq:E6}
27 = 16 + 10 + 1 \quad \text{under SO(10)} .
\end{equation}
The matter content needed by the $16_i + 10_i$ embedding can therefore be provided in the context of $E_6$ by three matter $27_i$, $i=1,2,3$, and the $16_H$ and $10_H$ needed to accommodate the Higgs fields can also be provided by a single Higgs $27_H$. All Yukawas can then in principle follow from the single $E_6$ interaction
\begin{equation}
\label{eq:E6Y}
\lambda_{ij} 27_i 27_j 27_H .
\end{equation}
We postpone the analysis of this promising possibility to further study. 

\section{Some solutions to the $\mu$-problem}
\label{sec:mu}

In this Section, we discuss a few approaches to the $\mu$-problem in the context of tree level gauge mediation. Let us remind what the $\mu$-problem is. Any supersymmetric extension of the SM must contain two Higgs doublet chiral superfields $\hat h_u$, $\hat h_d$, with hypercharges $\pm 1/2$, within  the light spectrum $Q$. Moreover, the lagrangian must contain a mass term for their Higgsino (fermion) components, $\mu\, \tilde h_u \tilde h_d$. It should also contain a corresponding term for the scalar components, $B_\mu h_u h_d$, where $B_\mu$ is a dimension two parameter. The Higgsino mass $\mu$ is constrained to be in the window $100\GeV \lesssim \mu \lesssim \TeV$ by the present bounds on chargino masses and by naturalness considerations. This coincides with the window for the supersymmetry breaking scale in the observable sector, $100\GeV \lesssim \tilde m \lesssim \TeV$. It is then tempting try to establish a connection between these two a priori independent scales, in such a way that $\mu\to 0$ when $\tilde m\to 0$, thus making the coincidence of the two scales not accidental. This is the $\mu$-problem. In the absence of such a connection, there would be no reason why $\mu$ should not be of the order of a much larger, supersymmetry conserving scale such as the GUT or the Planck scale. Or, if a symmetry or some other independent principle suppressed $\mu$, there would be no reason why $\mu$ should not be much smaller. 

Fermion mass terms such as $\mu\, \tilde h_u \tilde h_d$ belong to the list of possible soft supersymmetry breaking mass terms~\cite{softterms}. The reason why they are usually omitted from the MSSM effective soft supersymmetry breaking lagrangian is that they can be always reabsorbed in the superpotential (through appropriate additions to the scalar soft lagrangian). Moreover, most models of supersymmetry breaking, including the ones we are considering, do not generate such supersymmetry breaking fermion mass terms. We can then assume that the Higgsino mass term arises from a corresponding term in the superpotential. The problem is then to relate the coefficient of that (supersymmetric) superpotential term, $\mu\, \hat h_u \hat h_d$, to the supersymmetry breaking scale in the observable sector, which in our case is given by $\tilde m \sim |F_0|/M_V$. We discuss in the following three possible connections. One is peculiar of tree level gauge mediation, the other two have been considered in other contexts, but have specific implementations in tree level gauge mediation.  We classify them according to the dimension $D$ of the SO(10) operator from which the $\mu$ term arises.  Note that we are not addressing the origin of the smallness of $\tilde m$ and $\mu$ compared to the Plank scale, just their connection. The three options we consider are:
\begin{description}
\item[$D= 3$:] $\mu$ comes from the operator $\mu\, \hat h_u \hat h_d \subset W$. It is the supersymmetry breaking scale to be derived from $\mu$, and not viceversa: $F_0 \sim \mu M$, where $M = \ord{M_V}$,  and $\tilde m \sim F_0/M \sim \mu$. 
\item[$D=4$:] $\mu$ comes from the operator $\lambda S \hat h_u \hat h_d \subset W$. The light SM singlet $S$ gets a vev from a potential whose only scale is $\tilde m$, so that $\mu\sim \lambda \vev{S} \sim \tilde m$. 
\item[$D=5$:] $\mu$ comes from the operator $a (Z^{\dagger}/M) \hat h_u \hat h_d \subset K$, so that $\mu = a F_0/M$.
\end{description}
Let us discuss each of those possibilities in turn. 

\subsection{$D=3$} 

Such a possibility was anticipated in~\cite{NRZ}, where however no concrete implementation was given. Let us consider the $16_i+10_i$ embedding. As discussed in Section~\ref{sec:1610}, $\hat h_u$ is a superposition of the ``up Higgs-type'' components in the $\overline{16}$'s and $10$'s (with $R_P = 1$) in the model. Analogously, $\hat h_d$ will be a superposition of the ``down Higgs-type'' components in the ($R_P=1$) $16 $'s and $10$'s. The only possible $D=3$ origin of the $\mu$-term in the context of the full SO(10) theory are then $\ord{\text{TeV}}$ mass terms for the above $\overline{16}$'s, $16$'s, and $10$'s. As said, we do not address the origin of such a small parameter in the superpotential, as we do not address here the smallness of the supersymmetry breaking scale. The latter can for example be explained by a dynamical mechanism. We want however to relate such mass parameters, in particular  the coefficient of a $\overline{16} 16$ mass term, to the supersymmetry breaking scale. This is actually pretty easy, as the tree level gauge mediation embedding we are considering provides all the necessary ingredients and the result arises from their simple combination. We have seen in fact that the model needs a $16$, $\overline{16}$ pair to get a vev in the SM singlet direction of the scalar component, in order to break SO(10) to the SM. Moreover, we have seen that an independent $16'$, $\overline{16}'$ pair is required to break supersymmetry trough the $F$-term vev of the SM singlet component in the $16'$. The simplest way to achieve such a pattern is through a superpotential like
\begin{equation}
\label{eq:Wvev}
W_1 = \lambda_1 Z(\overline{16} 16 -M^2) + m 16' \overline{16} + \lambda_2 X 16 \overline{16}' ,
\end{equation}
where $X$, $Z$ are SO(10) singlets and $M\sim \mgut$. This is a generalization of an example in~\cite{ADGR}. Finally, we have just reminded that the light Higgses may have a component in $16$, $16'$, $\overline{16}$, $\overline{16}'$. Let $\alpha'$ be the coefficient of the $h_d$ component in the $16'$ and $\alpha$ the coefficient of the $h_u$ component in the $\overline{16}$. Then a $\mu$ parameter is generated in the form
\begin{equation}
\label{eq:mu3}
\mu = \alpha' \alpha\, m 
\end{equation}
from the $m 16' \overline{16}$ term in \eq{Wvev}. The parameter $m$ is therefore required to be in the window $100\GeV/(\alpha' \alpha) \lesssim m \lesssim \TeV/(\alpha' \alpha)$. In the limit $\mu = 0$, supersymmetry is unbroken and $16$, $\overline{16}$ acquire a vev that can be rotated in the SM singlet component $\vev{1^{16}} = \big\langle1^{\overline{16}}\big\rangle = M$. A non-vanishing $\mu$, on the other hand, triggers supersymmetry breaking and induces an $F$-term vev for the singlet component of the $16'$, $ \big\langle1^{16'} \big\rangle = F\theta^2$, with $F = m M$. We therefore have
\begin{equation}
\label{eq:connection3}
\tilde m \sim \frac{F}{M} = m = \frac{\mu}{\alpha' \alpha} ,
\end{equation}
providing the desired connection between $\mu$ and the supersymmetry breaking scale. Tree level gauge mediation plays a crucial role not only in providing the ingredients (and no need to stir) but also because it is the very SO(10) structure providing the heavy vector messengers to relate in a single irreducible representation (the 16') supersymmetry breaking (the $F$-term vev of its SM singlet component) and the down Higgs entering the $\mu$-term (the lepton doublet-type component of the 16'). In the Appendix~\ref{sec:W} we provide an existence proof of a (perturbative) superpotential that i) implements the mechanism above, thus breaking supersymmetry and SO(10) to SU(5), ii) further breaks SU(5) to the SM, iii) makes all the fields that are not part of the MSSM spectrum heavy, in particular achieves doublet-triplet splitting. 

\subsection{$D=4$} 

This is an implementation of the NMSSM solution of the $\mu$-problem~(see e.g. \cite{ellwanger} and references therein). As we will see, the implementation of such a solution in the context of tree level gauge mediation avoids some of the problems met in ordinary gauge mediation. 

In order to implement the NMSSM solution of the $\mu$-problem, an explicit term $\mu\, \hat h_u \hat h_d$ should be forbidden, for example by a symmetry; the light fields $Q$ should include a SM singlet $S$, coupling to the Higgses through the superpotential interaction $\lambda \hat S \hat h_u \hat h_d$; and $S$ should develop a non-zero vev. A $\mu$ parameter will then be generated, $\mu = \lambda \vev{S}$. In the absence of terms linear or quadratic in $\hat S$ in the superpotential, the scale of a vev for $S$ can only be provided by the supersymmetry breaking terms in the soft lagrangian, $\vev{S} \sim  \tilde m$, in which case $\mu = \lambda \vev{S} \sim \lambda \tilde m$, as desired. 

In order to generate a non-zero vev for $S$, one would like to have a negative soft mass for $S$ at the weak scale, along with a stabilization mechanism for large values of the fields. In ordinary gauge mediation this is not easy to achieve. While the stabilization can be simply provided by a $S^3$ term in $W$, as in the NMSSM (or by a quartic term in $Z'$ extensions of the MSSM~\cite{Langacker:1999hs}), the soft mass term of $S$ vanishes at the messenger scale because $S$ is typically a complete gauge singlet. A non-vanishing negative mass term is generated by the RGE running but it is typically too small. Another problem is that the Higgs spectrum can turn out to be non-viable~\cite{GR144}. A sizable soft mass can still be generated by coupling $S$ to additional heavy fields. Such possibilities can be implemented in our setup by promoting $S$ to an SO(10) singlet and coupling it to the Higgses through a $S\, \overline{16}\, 16$ or a $S\, 10\, 10$ coupling to the SO(10) representations containing (a component of) the Higgs fields. 

Tree level gauge mediation offers a different avenue. A sizable, negative soft mass term for $S$ can in fact be generated by embedding $S$ in a $\overline{16}$ of SO(10) (this is the only choice within the fields in Table~\ref{tab:X}). On the other hand, the stabilization of the potential for $S$ is not straightforward. A sizable $S^3$ term is not expected to arise, as it should involve a SO(10) operator with three $\overline{16}$. However, the $S^3$ term can be replaced by a term involving a second light singlet $N$,
\begin{equation}
\label{eq:NMSSM}
W = \lambda \hat S \hat h_u \hat h_d + \kappa \hat S^2 \hat N .
\end{equation}
The latter can come from a $\overline{16}^2 126$ coupling, if $N$ is in the 126 singlet, or from a $\overline{16}^2 16_1 16_2/\Lambda$ coupling, where $N$ is the $16_1$ singlet and $16_2$ gets a vev.

The scalar potential for $V(h_u,h_d,S,N)$ can be written as
\begin{equation}
\label{eq:VNMSSM}
V = V_\text{MSSM} + |\kappa S^2|^2 + m_S^2 |S|^2 + |\lambda h_u h_d + 2 \kappa S N|^2 + M_N^2 |N|^2 ,
\end{equation}
where $V_\text{MSSM}$ is the MSSM scalar potential with $\mu\to \lambda S$, $m^2_S = -\tilde m^2$, and $m^2_N = 2\tilde m^2$ or $\tilde m^2$ depending on whether $N$ comes from a 126 or a 16. We have neglected the $A$-terms, which play a role in explicitly breaking $R$-symmetries that could lead to massless states. The potential above has a minimum with a sizable $\vev{S}$, and a $\mu$ parameter whose size is controlled by $\lambda$. 

\subsection{$D=5$} 

Finally, let us consider the possibility to generate the $\mu$ parameter through a $D=5$ correction to the K\"ahler in the form $a (Z^\dagger/M) \hat h_u \hat h_d$, as in the Giudice Masiero mechanism~\cite{GiudiceMasiero}. The $F$-term vev $|F_0|$ of $Z$ would give in this case $\mu = a |F_0|/M$. 

We show first that the operator above cannot arise at the tree level from integrating out heavy vector or chiral superfields. The corrections to the K\"ahler obtained by integrating out heavy vector superfields are given in \eqs{Keff2}. All terms are at least of second order in $1/M_V$ and no trilinear term is present. Moreover, no sizable trilinear term can be obtained through the vev of  $\Phi'$, as by definition the scalar components of $\Phi'$ do not get a vev (and an $F$-term vev would give an additional $F_0/M_V$ suppression). A similar conclusion can be obtained for the corrections one obtains by integrating out chiral superfields $\Phi^h_i$ with mass $M\gg \sqrt{|F_0|}$. We have seen in Section~\ref{sec:tree} that the equations of motion allow to express $\Phi^h_i$ in terms of the light fields as in \eq{Phih}. Since $W_3$ contains terms at least trilinear in the fields, the expression for $\Phi^h_i$ is at least quadratic in the light fields. When plugging \eq{Phih} in the canonical K\"ahler for $\Phi^h_i$ one gets again terms that contain at least four light fields, with none of them getting a vev in the scalar component. Therefore, no operator $Z^\dagger \hat h_u \hat h_d$ can be generated at the tree level by integrating out heavy fields. 

Let us now consider the possibility that the $D=5$ operator above is obtained at the one loop level. This possibility raises two issues. First, $\mu$ would be suppressed compared to, say, the stop mass $\tilde m_t$ by a loop factor $\ord{10^{-2}}$. As for the case of gaugino masses vs sfermion masses, such a large hierarchy would lead to sfermions beyond the reach of the LHC and a significant fine-tuning. However, as we will see, this problem can be overcome in the same way as for the gaugino masses. We will see in fact in an explicit model that $\mu$ and $M_{1/2}$ get a similar enhancement factor. The second issue is the well known $\mu$-$B_\mu$ problem. $B_\mu$ is a dimension two parameter generated, as $\mu$, at the one loop level. Therefore, we expect an order of magnitude separation between $\sqrt{B_\mu}$ and $\mu$: $\sqrt{B_\mu}/\mu \sim 4\pi$. This is however tolerable in a scheme in which $\tilde m_t \sim \sqrt{B_\mu} \sim 4\pi \mu \sim 4\pi M_{1/2}$, with $\tilde m_t \sim \sqrt{B_\mu} \sim \TeV$ and $\mu\sim M_{1/2} \sim 100\GeV$. The explicit model will show that the above pattern can be achieved in the large $\tan\beta$ regime. In turn, the large $\tan\beta$ regime raises a new issue. The minimization of the MSSM potential shows in fact that large $\tan\beta$ corresponds to small $B_\mu/(m^2_{h_u} + m^2_{h_d}+2|\mu|^2)$, while in the situation we want to reproduce, $\tilde m_t \sim \sqrt{B_\mu}$, we expect $B_\mu/(m^2_{h_u} + m^2_{h_d}+2|\mu|^2) \sim 1$. In order to make $\tan\beta$ large we therefore need to cancel the contribution to $B_\mu$ we get at one loop with an additional contribution, at least in the specific example we consider. Such a cancellation may not be required in different implementations of the one-loop $D=5$ origin of the $\mu$ parameter. That is why we believe it is worth illustrating the example below despite the cancellation that needs to be invoked. 

Let us consider as before a model involving the following $R_P = 1$ fields: $16$, $\overline{16}$, $16'$, $\overline{16}'$, $10$, with $\vev{1^{16}} = \big\langle1^{\overline{16}}\big\rangle = M$, $\big\langle1^{16'}\big\rangle = F\theta^2$, $\big\langle1^{\overline{16}'} \big\rangle = 0$. Let us denote the coefficients of the $h_u$ and $h_d$ components in the above SO(10) representations as follows: $16 \supset s_d\alpha_d h_d$, $16' \supset s_d\alpha'_d h_d$, $10 \supset c_d h_d$, $\overline{16} \supset s_u\alpha_u h_u$, $\overline{16}' \supset s_u\alpha'_u h_u$, $10 \supset c_u h_u$, where $|\alpha_d|^2 + |\alpha'_d|^2 = 1$, $|\alpha_u|^2 + |\alpha'_u|^2 = 1$, $c_d = \cos\theta_d$, $s_d = \sin\theta_d$, etc. The notation is in agreement with the definition of $\theta_u$, $\theta_d$ in Section~\ref{sec:1610}. The $\mu$ and $B_\mu$ parameters, as the gaugino masses, get a vector and a chiral one-loop contribution, see eqs.~(\ref{eq:dKloop},\ref{eq:dKloop2},\ref{eq:mu},\ref{eq:Bmu}). The vector contribution turns out to be
\globallabel{eq:muvec}
\begin{align}
|(\mu)_V| & = \frac{3}{2} \frac{g^2}{(4\pi)^2} s_u s_d |\alpha'_d\alpha_u| \fracwithdelims{|}{|}{F}{M} \mytag \\
(B_\mu)_V & = \frac{3}{4} \frac{g^2}{(4\pi)^2} s_u s_d |\alpha'_d\alpha_u| \fracwithdelims{|}{|}{F}{M}^2 \mytag .
\end{align}
As in the case of gaugino masses, the vector contribution to $\mu$ is suppressed with respect to the sfermion masses by a full loop factor. We therefore need a larger chiral contribution in order to reduce the hierarchy between $\mu$ and $\tilde m_t$. Let us then consider the one-loop chiral contribution associated to the superpotential 
\begin{equation}
\label{eq:WPQ}
h_{ij} 16_i 10_j 16 + h'_{ij} 16_i 10_j 16'.
\end{equation}
That is easily found to be vanishing because of a PQ symmetry of the superpotential. Such a PQ symmetry can however be broken by adding a term 
\begin{equation}
\label{eq:WPQ2}
\frac{M^1_{ij}}{2} 1^{16}_i 1^{16}_j
\end{equation}
to the above superpotential, coming for example from the non-renormalizable SO(10) operator $(\alpha_{ij}/\Lambda) (\overline{16} 16_i)(\overline{16} 16_j)$ after $\overline{16}$ gets its vev (note that $\Lambda\gg M$ would give $M^1_{ij} \ll M$). En passant, the singlet mass term in \eq{WPQ2} is nothing but the right-handed neutrino Majorana mass term entering the see-saw formula for light neutrino masses. Note however that no light neutrino mass is generated here, as the light lepton doublets do not have Yukawa interactions with the ``right-handed neutrinos'', $1^{16}_i$. Once the PQ symmetry is broken by the mass term in \eq{WPQ2}, the $\mu$ and $B_\mu$ parameters get a chiral one-loop contribution given by 
\globallabel{eq:muchi}
\begin{align}
|(\mu)_\Phi| & = \frac{\lambda_t\lambda_b}{(4\pi)^2} f\left(
\frac{\sqrt{(M^1M^{1*})_{33}}}{|h_{33} M|} 
\right) \frac{|M^1_{33}|}{\sqrt{(M^1M^{1*})_{33}}} \fracwithdelims{|}{|}{h'_{33} F}{h_{33}M} \mytag \\
(B_\mu)_\Phi & = \frac{\lambda_t\lambda_b}{(4\pi)^2} g\left(
\frac{\sqrt{(M^1M^{1*})_{33}}}{|h_{33} M|} 
\right) \frac{|M^1_{33}|}{\sqrt{(M^1M^{1*})_{33}}} \fracwithdelims{|}{|}{h'_{33} F}{h_{33}M}^2 , \mytag
\end{align}
where $\lambda_t$, $\lambda_b$ are the top and bottom Yukawa couplings respectively and the functions $f$, $g$ are given by
\begin{equation}
\label{eq:fg}
f(x) = \frac{1-x^2+x^2\log x^2}{(x^2-1)^2} x, \qquad
g(x) = \frac{x^4 -2x^2\log x^2 -1}{(x^2-1)^3} x .
\end{equation}
We have assumed the Yukawa couplings $h_{ij}$, $h'_{ij}$ to be hierarchical in the basis in which the down Yukawa matrix is diagonal. 

We can see from \eq{muchi} that the one loop chiral contribution to $\mu$ is comparable to the corresponding contribution to $M_{1/2}$ if i) $\lambda_b \sim 1$, which corresponds to the large $\tan\beta$ regime (remember that the bottom mass is given by $m_b = \lambda_b \cos\beta v$, where $v=174\GeV$); ii) $|h'_{33}/h_3| \gtrsim  |h'_{ii}/h_i|$, $i=1,2$; iii) $|M_{33}| \gtrsim |M_{3i}|$; iv) $|h_{33}M| \sim |M_{33}|$. If the above conditions are satisfied, $\mu \sim M_{1/2}$ and both parameters can easily be enhanced, as explained in Section~\ref{sec:gauchi}, for example because $|h'_{33}/h_{33}|\gg 1$. The only non-trivial condition is the large $\tan\beta$ one. Remember in fact that $\tan\beta$ is determined by $B_\mu$ through the minimization of the MSSM potential, which gives 
\begin{equation}
\label{eq:BmuMSSM}
\sin 2\beta = \left.\frac{2B_\mu}{m^2_{h_u}+m^2_{h_d} + 2|\mu|^2} \right|_{M_Z}.
\end{equation}
Therefore large $\tan\beta$, i.e.\ small $\sin 2\beta$, requires a small $B_\mu$. This is in contrast with the  situation we want to reproduce, $\tilde m_t \sim \sqrt{B_\mu}$. The RGE evolution of $B_\mu$ from the scale at which it is generated ($|h_{33}M|$) down to the electroweak scale can reduce the value of $B_\mu$ but not enough to make it as small as we need. A significant RGE contribution would in fact require $M_{1/2}\gtrsim \tilde m_t$, in contrast with the $\tilde m_t \sim 4\pi M_{1/2}$ we are trying to reproduce. We are then forced to invoke a cancellation between the one-loop contribution to $B_\mu$ in eq.~(\ref{eq:muchi}b) and an additional contribution. For example, a tree level contribution to $B_\mu$ can be obtained as in Appendix B or in~\cite{ADGR}. 

\section{Conclusions}
\label{sec:conclusions}

In this paper we have considered what may be regarded as one of the simplest ways to communicate supersymmetry breaking from a hidden to the observable sector, through the tree level, renormalizable exchange of superheavy gauge (GUT) messengers, and we have studied the general properties of such a tree level gauge mediation (TGM) scheme. 

We have first of all obtained the general structure of the tree-level soft terms arising from a supersymmetry breaking source that is part of a non-trivial gauge (GUT) multiplet. This is most conveniently done in the effective theory in which the heavy vector superfields associated to the broken generators are integrated out at the tree level (en passant, in Appendix A, we summarized the procedure to integrate out vector superfields and addressed a few minor issues, such as the generalization to the non-abelian case and the role of gauge invariance in a consistent supersymmetric generalization of the expansion in the number of derivatives). The scalar soft terms then obtain the two contributions in \eq{mm}, corresponding to the two diagrams on the right-hand side of  \Fig{diagram}. Only the first contribution is relevant for scalars that are not in the same gauge multiplet as the scalar partner of the Goldstino (or have not the same SM quantum numbers as some of the GUT generators). Because of the tracelessness condition, such a contribution gives both positive and negative soft masses. This potential phenomenological problem, which has long been considered as an obstacle to tree level supersymmetry breaking, is automatically solved in the models we consider because the fields getting a $\ord{\text{TeV}}$ negative soft mass also get an $\ord{M_\text{GUT}}$ positive, supersymmetric mass. 

Gaugino masses do not arise at the tree level, but can be generated at the one-loop level, as in ordinary gauge mediation. They receive two contributions, from loops involving heavy vector or chiral superfields. The loop factor suppression of gaugino compared to sfermion masses must be at least partially compensated if the sfermions are to be within the LHC reach and the split-supersymmetry regime is to be avoided. We calculated in full generality the vector and chiral contributions to gaugino masses corresponding to the diagrams in Fig.~\ref{fig:mg}. We have seen that the vector contribution is always suppressed by a full loop factor and is typically subdominant (often vanishing). On the other hand, the chiral contribution is typically larger. We listed four potential enhancement factors that can (do) compensate, at least partially, the loop suppression: a larger number of (chiral) messengers contributing to gaugino masses than (vector) messengers contributing to sfermion masses; group theoretical factors that in practice turn out to enhance gaugino masses; the fact that sfermion masses are suppressed by all the vevs with non-vanishing gauge coupling to the vector messengers, while gaugino masses are suppressed only by the vevs that are related to supersymmetry breaking through superpotential interactions; ratios of Yukawa couplings appearing in the expression for the gaugino masses. In minimal models the first two factors partially compensate the $\ord{10^{-2}}$ loop factor, reducing it to the level of a tolerable (and possibly necessary) one order of magnitude hierarchy between gauginos and sfermions. The last two factors are more model-dependent but can give rise to larger enhancements. 

The general analysis of the TGM scheme allowed us to define the guidelines to obtain  phenomenologically viable models from the general formalism and to identify the assumptions underlying the peculiar predictions one obtains. Clear model building guidelines emerge, identifying SO(10) and $E_6$ as the ``minimal'' grand-unified groups, while SU(5) is found not to have the necessary structure (rank $\geq 5$) to realize the TGM scheme. The SO(10) possibility turns out to be quite appealing. It turns out in fact that the SO(10) breaking vevs of a $16+\overline{16}$, important to break SO(10) to the SM, typically make heavy precisely the components of the SO(10) representations that need to be made heavy because of their negative soft supersymmetry breaking masses. This reinforces the logical consistency of the TGM framework. 

In SO(10), the tree level sfermion soft masses turn out to be proportional to their charges under the SU(5)-invariant SO(10) generator $X$. We find two possible embeddings of the MSSM superfields into SO(10) representations, depending on whether $F^\dagger_0 X F_0$ is positive or negative. In the first case, $F^\dagger_0 X F_0>0$, the three MSSM families are embedded in three $16_i$ and three $10_i$, $i=1,2,3$. The quark doublets, the up quark singlets, and the lepton singlets, unified in 10's of SU(5), are embedded in the $16_i$'s, while the lepton doublet and down quark singlets, unified in $\overline{5}$'s of SU(5), are embedded in the $10_i$'s. They all get positive soft masses. The spare components in the $16_i$'s and $10_i$'s get superheavy, positive, supersymmetric mass terms (and TeV scale negative soft masses).  In the second case, $F^\dagger_0 X F_0<0$, the three MSSM families are embedded in three $16_i$ and three $45_i$, $i=1,2,3$. The MSSM fields in 10's of SU(5) are embedded in the $45_i$'s, while the ones unified in $\overline{5}$'s of SU(5), are embedded in the $16_i$'s. As before, they all get positive soft masses. In both cases the chiral content of the theory is still given by three 16 of SO(10). An important property of the TGM soft terms is that they turn out to be family universal, thus solving the supersymmetric flavor problem. This property only depends on the hypothesis that the three MSSM families are embedded in the same SO(10) representations. Mixed embeddings, in which the MSSM fields are superpositions of fields in inequivalent SO(10) representations, are also possible, but can spoil the flavor universality property. Each of the two possible flavor-universal embeddings leads to specific and peculiar predictions for the soft masses at the GUT scale: $m^2_{\overline{5}} = 2 m^2_{10}$ in the $F_0^\dagger X F^{\phantom{\dagger}}_0>0$ case and $m^2_{\overline{5}} = (3/4) m^2_{10}$ in the $F_0^\dagger X F^{\phantom{\dagger}}_0<0$ case, where $m^2_{\overline{5}}$ and $m^2_{10}$ are common and family-independent soft masses for the fields in the $\overline{5}$ and $10$ of SU(5) respectively. The latter predictions are only based on i) the use of the ``minimal'' unified gauge group SO(10), ii) the embedding of the MSSM families in the SO(10) representations with dimension $d<120$ not containing the Goldstino, and iii) the absence of mixed embeddings to automatically preserve flavor-universality. The predictions on the ratios $m_{\overline{5}}/m_{10}$ in \eq{prediction} are determined by group theory factors and are peculiar enough to make a possible experimental test at the LHC a strong hint for tree level gauge mediation. The embedding into three $16_i+10_i$'s has the advantage that the large top Yukawa coupling can be accounted for by a renormalizable superpotential interaction involving only low-dimensional ($d\leq 16$) representations for the chiral superfields. In the $16_i+45_i$ case, a $d=210$ representation of SO(10) must be used to reproduce the top Yukawa coupling at the renormalizable level. 

The $E_6$ option is also quite appealing, as the matter superfield content of the $16_i+10_i$ embedding is precisely the one obtained from three fundamentals $27_i$ of $E_6$. The latter decompose in fact as $27_i = 16_i + 10_i + 1_i$ under SO(10). We have postponed the investigation of this promising possibility to further study. 

Finally, we have illustrated three possible approaches to the $\mu$-problem in TGM, which we classify according to the dimension $D$ of the SO(10) operator from which the $\mu$-term arises. The $D=3$ option provides a new approach to the $\mu$-problem, peculiar of TGM. The idea is that the supersymmetry breaking scale turns out to coincide with the $\mu$ scale because supersymmetry is triggered by the same $D=3$ SO(10) operator from which the $\mu$-term arises. We have provided an explicit realization of such a possibility in Appendix B. The superpotential shown there also achieves supersymmetry breaking, SO(10) breaking to the SM, and ensures that only the MSSM fields survive below the breaking scale (in particular it provides doublet-triplet splitting). While it is not meant to be simple or realistic, that superpotential represents a useful existence proof. The $D=4$ option is nothing but the NMSSM solution of the $\mu$ problem, in which the $\mu$-term is obtained from the vev of a SM singlet superfield stabilized at the supersymmetry breaking scale. We pointed out that the above singlet can easily get a sizable, predictable, negative soft mass term in TGM. This makes giving a vev to the singlet easier than in ordinary gauge mediation (where its soft mass usually vanishes before RGE running), provided that the singlet potential can be made stable. The $D=5$ option is nothing but the Giudice-Masiero mechanism realized at the loop level, as in gauge mediation. The consequent loop hierarchy between the $\mu$-term and the sfermion masses can be reduced exactly as for the gaugino masses. We provided an explicit example, which however needs an extra contribution to the $B_\mu$ parameter in order to give rise to the necessary large $\tan\beta$. 

\section*{Acknowledgments}

We thank Andrea Brignole and Claudio Scrucca for useful discussions and acknowledge partial support from the RTN European Program ``UniverseNet'' (MRTN-CT-2006-035863). We thank the Galileo Galilei Institute for Theoretical Physics, where part of this work was done, for the hospitality. 

\appendix

\section{Integrating out vector superfields}
\label{sec:eom}

In this Appendix, after a few general comments, we write the effective theory one obtains by integrating heavy vector superfields at the tree level and in unitary gauge in a generic, non-abelian, $N=1$ globally supersymmetric theory with renormalizable K\"ahler $K$ and gauge-kinetic function (the superpotential $W$ is allowed to be non-renormalizable). The general prescription has been studied in~\cite{scrucca,refsthereintogether,ISlectures}. In particular, it has been shown in~\cite{scrucca} that the usual expansion in the number of derivatives $n_\partial$ can be made consistent with supersymmetry by generalizing $n_\partial$ to the parameter
\begin{equation}
\label{eq:n}
n = n_\partial + \frac{1}{2} n_\psi + n_F,
\end{equation}
where $n_\psi/2$ is the number of fermion bilinears and $n_F$ the number of auxiliary fields from chiral superfields. With such a definition, a chiral superfield $\Phi$ has $n=0$ and $d\theta$ integrations and supercovariant derivatives have $n=1/2$. Such an expansion makes sense when supersymmetry breaking takes place at a scale much smaller than the heavy superfield mass $M$ and in particular when the $F$-terms and fermion bilinears from heavy superfields being integrated out are much smaller than $M$. 

In the presence of vector superfields one should further assume that the $D$-terms and gaugino bilinears are small and should generalize \eq{n} to account for the number $n_\lambda$ of gauginos and the number $n_D$ of vector auxiliary fields. We claim that the correct generalization is 
\begin{equation}
\label{eq:n2}
n = n_\partial + \frac{1}{2} n_\psi + n_F + \frac{3}{2} n_\lambda + 2 n_D,
\end{equation}
according to which a vector superfield $V$ has $n=0$. Note that the double weight of $D$-terms compared to $F$-terms is consistent with \eq{Dh}. With such a definition, the initial lagrangian has $n=2$, except for the gauge kinetic term, which has $n=4$. Chiral and vector superfields can then be integrated out at the tree level by using the supersymmetric equations of motion
\begin{equation}
\label{eq:susyeom}
\frac{\partial W}{\partial \Phi} = 0 \quad \text{and} \quad 
\frac{\partial K}{\partial V} = 0 
\end{equation}
up to terms with $n\geq 3$ when integrating out chiral superfields and $n \geq 4$ when integrating vector superfields (with the missing terms originating from the gauge-kinetic term having $n\geq 6$). 

From a physical point of view, we are interested not only in the expansion in $n$ but also, and especially, in the expansion in the power $m$ of $1/M$. It is therefore important then to remark that using \eqs{susyeom} amounts to neglecting terms with $m \geq 3$ when integrating chiral superfields and $m\geq 6$ when integrating out vector superfields. 

\interskip

We are now ready to present out results on the effective theory obtained integrating out the heavy vector superfields in a generic supersymmetric gauge theory as above. We are interested in operators with dimension up to 6 ($m \leq 2$) in the effective theory. We can then use the equation $\partial K/\partial V$. Neglecting higher orders in $m$, the latter equation can be rewritten as
\begin{equation}
\label{eq:Vh}
V_a^h (M_V^2)^{\phantom{h}}_{ab} = - \frac{1}{2} \frac{\partial K_2}{\partial V^h_b} (\Phi^\prime,V^l) , 
\end{equation}
where $\Phi'$ is defined in \eq{goldstones}, $K_2(\Phi',V) = \Phi'^\dagger e^{2gV}\Phi'$, the indices run over the broken generators, and $M^2_V$ is a function of the light vector superfields:
\begin{equation}
\label{eq:MMVexp}
\begin{gathered}
(M_V^2)_{ab}  =  \frac{1}{2}
\frac{\partial^2}{\partial V^h_a \partial V^h_b}
\left. \left( \phi^\dagger_0 e^{2gV} \phi_0 \right) \right|_{V^h = 0} = 
(M_{V0}^2)_{ab} +  (M^2_{V2})_{ab} \\
(M^2_{V0})_{ab}   = g^2 \phi_0^* \{ T_a^h, T^h_b \} \phi_0 \\[1mm]
(M^2_{V2})_{ab}  =  \frac{g^4}{3} \phi_0^* T_a^h V^l V^l T^h_b \phi_0 + (a \leftrightarrow b)  .
\end{gathered}
\end{equation}

In order to solve \eq{Vh} for $V^h_a$, we need to invert the field-dependent matrix $M^2_V$. In the Wess-Zumino gauge for the light vector superfields, we get
\begin{equation}
\label{eq:MMinverse}
(M^2_{V})^{-1}_{ab} = (M^2_{V0})^{-1}_{ab} - (M^2_{V0})^{-1}_{ac} (M^2_{V2})_{cd} (M^2_{V0})^{-1}_{db} .
\end{equation}
The effective contribution to the K\"ahler potential is 
\begin{equation}
\label{eq:Keff1}
K_\text{eff} = -(M^2_V)_{ab} V^h_a V^h_b = K_0 + K_1 + K_2 ,
\end{equation}
where 
\globallabel{eq:Keff2}
\begin{align}
\delta K^0_\text{eff} & = - g^2 (M_{V0}^2)^{-1}_{ab}  (\Phi'^\dagger T^h_a \Phi') (\Phi'^\dagger T^h_b \Phi') \mytag \\[1.5mm]
\delta K^1_\text{eff} & = - 2 g^3 (M_{V0}^2)^{-1}_{ab} (\Phi'^\dagger T^h_a \Phi') (\Phi'^\dagger \{V^l,T^h_b\} \Phi') \mytag \\
\delta K^2_\text{eff} & = -  \frac{4}{3} g^4 (M_{V0}^2)^{-1}_{ab} (\Phi'^\dagger T^h_a \Phi') \Phi'^\dagger (T^h_b V^l V^l + V^l T^h_b V^l + V^l V^l T^h_b) \Phi' \mytag \\
& - g^4 (M_{V0}^2)^{-1}_{ab} \left( (\Phi'^\dagger \{ T^h_a,V^l \} \Phi') (\Phi'^\dagger \{ T^h_b,V^l \} \Phi') + \frac{1}{3} (\Phi'^\dagger [ T^h_a,V^l ] \Phi') (\Phi'^\dagger [ T^h_b,V^l ] \Phi') \right). \notag
\end{align}

In recovering eq.~(\ref{eq:Keff2}c) we have used the identity
\begin{equation}
\label{eq:identity}
f_{\alpha ab}(M^2_{V0})_{bc} = -f_{\alpha cb} (M^2_{V0})_{ba} ,
\end{equation}
where $f_{abc}$ are the structure constants of the gauge group, the latin indices refer to broken generators and the greek one refers to an unbroken one. 

\interskip

We are interested to soft supersymmetry breaking terms arising from \eqs{Keff2} when some of the auxiliary fields get a vev. The relevant terms should contain up to two $F$-terms and one $D$-term (\eq{Dh}). The only relevant terms are therefore those in (\ref{eq:Keff2}a). 

\section{Supersymmetry breaking, SO(10) breaking, $\mu$, and doublet-triplet splitting}
\label{sec:W}

In this Appendix we provide an example of a superpotential achieving supersymmetry breaking, SO(10) breaking to the SM, ensuring that below the scale of this breaking only the MSSM fields survive (in particular providing doublet-triplet splitting) and solving of the $\mu$-problem. We do not aim at being simple or realistic, we just aim at providing an existence proof. We include in this example representations with dimension $d>120$. It would be interesting to obtain a dynamical supersymmetry breaking, in particular the $F$-term vev of a 16 of SO(10). 

SO(10) will broken to the SM at a scale $M \sim M_\text{GUT}$. Below this scale only the MSSM fields survive, in particular the Higgs triplets are made heavy via a generalization of the Dimopoulos-Wilczek mechanism~\cite{DW,CFLR}. The $\mu$-term is present in the theory in the form of a $D=3$ operator present at the GUT scale and triggers supersymmetry breaking. $B _\mu$ is generated at the tree-level and turns out to be of the same order as the sfermion masses. 

\subsection{The superpotential}

The superpotential we use is 
\begin{equation}
\label{eq:fullW}
W = W_{Y} + W_1 + W_2 +W_3 + W_4 , 
\end{equation}
where
\begin{equation}
\begin{aligned}
W_{Y} & = y_{ij} 16_i 16_j 10 + h_{ij} 16_i 10_j 16 + h_{ij}^\prime 16_i 10_j 16^\prime \\
W_1  & = \la{1} Z  (\Sb  16 -M ^2) + m \, \Sb \Sp + \la{2} X \, \Sbp 16 \\
W_2 & = \Sbpp (\la{3} 45 + \la{4} U) 16 + \Sb (\la{5} 45 + \la{6} \Up) \Spp + \MA \, 45 \, 45 + \la{7} 54 \, 45 \, 45 + \MS \, 54 \, 54 \\
W_3 & = \la{8} 16 \Sp 120 + \la{9} \Sb \, \Sbp 120 + \Mhun \, 120 \, 120  \\ 
W_4 & = \la{10} \Tp \, 45 \, 10 + \la{11} \Sb \, \Sbpp 10 + \MT \Tp \Tp  + \la{12} \Sb \, \Sbp 10 + \la{13} 16 \Spp 10 + \la{14} Z \, 10 \, 10.
\end{aligned}
\end{equation}
Here we denote the fields according to their SO(10) representation, except the SO(10) singlet fields $ Z,X,U, \Up$. The mass parameter $m$ is of the order of the TeV scale (we do not discuss the origin of such a small parameter here), 
while all other mass parameters are near the GUT scale  $$ \text{TeV} \sim m \ll M \sim \MA \sim \MS \sim \MT \sim \Mhun \sim M_{\rm GUT}. $$ 

Let us discuss the role of the different contributions to the superpotential and anticipate the vacuum structure and the spectrum. $W_1$ is responsible for supersymmetry breaking and the breaking of SO(10) to SU(5): as we are going to show below, this part of the superpotential generates $\ord{M_\text{GUT}}$ vevs for the scalar components of $16$ and $\Sb$ along the SU(5) singlet direction $\langle S \rangle \sim M + {\cal O}(m^2/M) $ and  $\langle \Sib \rangle \sim M + {\cal O}(m^2/M)$ and a supersymmetry breaking vev for the $F$-term component of $\Sp$  along the SU(5) singlet direction $\langle F_{\Sip} \rangle \sim m M $. It also provides small supersymmetry breaking vevs for the $F$-term component of $X$ $\langle F_X \rangle \sim m^2$ and  for the $D$-term of the vector superfield corresponding to the U(1)$_X$ generator of SO(10) $\langle D_X \rangle \sim M (\langle S \rangle - \langle \Sib \rangle) \sim m^2.$ This $D$-term vev will generate sfermion masses along the lines of Section \ref{sec:1610}. This superpotential appears in \eq{Wvev} and is a generalization of an example in~\cite{ADGR}. 

$W_Y$ contains the MSSM Yukawa couplings and provides supersymmetry breaking masses for heavy chiral superfields that will generate gaugino masses at 1-loop as in ordinary gauge mediation. The MSSM matter is embedded in both the $16_i$ and the $10_i$, as explained in Section~\ref{sec:1610}. The MSSM Higgs fields are linear combinations of different fields and have components in different representations, $$h_u \subset 10, \Sb \qquad  h_d \subset 10, 16, \Sp, 120.$$ Therefore the first term in $W_Y$ contains the up-type Yukawas, while the second and third terms provide down-type and charged lepton Yukawas. The second term also gives a large mass to the additional fields $5^{10}_i \subset 10_i$ and $\overline{5}^{16}_i \subset 16_i.$ The latter are also the only fields that couple to the $F$-term vev in the $\Sp$ and act as one-loop messengers of supersymmetry breaking. While this gives a subleading contribution to sfermion masses, it is the only source of gaugino masses in this model.\\

The role of $W_2$ is the breaking of SU(5) to the standard model gauge group. It provides a large vev for the $45$ along the $B-L$ direction $\langle 45_{B-L} \rangle \sim M$  as needed for the Dimopoulos-Wilzcek mechanism. Also $U, \Up$ and the SM singlet in the $54$ take large vevs. $W_3$ merely gives large masses to components in the $\Sp$ and $\Sbp$. Note that since the $120$ does not contain SU(5) singlets, the neutrino component in the $\Sp$ stays massless as it should, being the dominant component of the Goldstino superfield. $W_4$ takes care of the Higgs sector: it keeps the MSSM Higgs doublets light and gives a large mass to the corresponding triplets. Its last term provides the $B_\mu$ term because $Z$ gets a small supersymmetry breaking vev and both $H_u$ and $H_d$ have components in the $10$. The $\mu$ term is contained in $W_1$, because $H_d$ has a component also in the $\Sp$.

\subsection{The vacuum structure}

We are interested in a vacuum that does not break the SM gauge group. Thus only that part of the superpotential which involve SU(5) singlets is relevant for the determination of the ground state. We denote the singlets in $(16,\Sb,\Sp,\Sbp,\Spp,\Sbpp)$ by $(S,\Sib,\Sip,\Sibp,\Sipp,\Sibpp)$ (which is different than the notation used in the main text) and the singlets in the $45,54$ by $B,T,V$, where $B,T$ are the properly normalized fields corresponding to the $B-L$ and $T_{3R}$ generators in SO(10). The relevant part of the superpotential is

\begin{align}
W & =  \la{1} Z  (\Sib  S - M^2) + m \, \Sib \Sip + \la{2} X \, \Sibp S \notag \\[2mm]
& + \Sibpp \left( -\frac{\la{3}}{2} T + \frac{\la{3}}{2} \sqrt{\frac{3}{2}} B + \la{4} U\right) S + \Sib \left( -\frac{\la{5}}{2} T + \frac{\la{5}}{2} \sqrt{\frac{3}{2}} B + \la{6} U \right) \Sipp \\[-1mm]
& + \MA (B^2 + T^2) + \MS V^2 + \la{7} V \left( \frac{1}{2} \sqrt{\frac{3}{5}} T^2 - \frac{1}{\sqrt{15}} B^2 \right). \notag
\end{align}
The $F$-term and $D_X$-term equations show that SUSY is broken ($F_{\Sip} \neq 0$) and that all vevs are determined except $V,B,T$, for which there exist three solutions, all yielding $F_T = F_V = F_B =0$. This tree-level degeneracy is lifted by one-loop corrections which select the solution with $T=0, B\neq 0, V \neq 0$. One can check that the vevs are given by
\begin{gather}
\label{eq:vevs}
\Sip = \Sibp = \Sipp = \Sibpp = X = Z = T = 0  \notag \\[2mm]
S = M - \frac{m^2}{4 M} \left(\frac{1}{\lambda_1^2} - \frac{1}{50 g^2}\right ) \qquad \Sib = M - \frac{m^2}{4 M} \left(\frac{1}{\lambda_1^2} + \frac{1}{50 g^2}\right) \notag  \\
U = - \frac{3 \sqrt{5} }{2} \frac{\la{5}}{\la{6} \la{7}} \sqrt{\MA \MS} \qquad \Up =  - \frac{3 \sqrt{5} }{2} \frac{\la{3}}{\la{4} \la{7}} \sqrt{\MA \MS}  \\[1mm]
V = \frac{ \sqrt{15} \MA}{\la{7}} \qquad B =  \frac{\sqrt{30}}{\la{7}}  \sqrt{\MA \MS}  \notag \\
F_{\Sip} = - m M \qquad F_Z = \frac{m^2}{2 \la{1}}  \qquad D_X = - \frac{m^2}{10 g}. \notag 
\end{gather}   

\subsection{Spectrum and soft terms}
 
In order to identify the light (with respect to $M_\text{GUT}$) we can set $m=0$ and consider the supersymmetric limit. Most fields are at the GUT scale, with the light ones being the MSSM ones, the Goldstino superfield $\Sip$, and the right-handed neutrinos in the $16_i$, which can easily be made heavy through a non-renormalizable superpotential operator $(\overline{16} 16_i)(\overline{16} 16_j)$. The MSSM matter fields are embedded in the $10^{16}_i$ and in the $\overline{5}^{10}_i$, as desired. The Higgs doublets are embedded into the $\Sb,10$ and $10, 120, \Sp, 16$ according to 
\begin{equation}
\label{eq:higgsembedding}
\begin{gathered}
h_u  = \frac{1}{N_u} \left(\overline{L}_{\Sb} + 3 \sqrt{5} \frac{\la{5}}{\la{7} \la{13}} \frac{\sqrt{\MA \MS}}{M} \overline{L}_{10}\right) \\
h_d  = \frac{1}{N_d} \left(L_{10} - \frac{\la{12}}{\la{9}} L_{120} + 2 \frac{\la{12}}{\la{8} \la{9}} \frac{\Mhun}{M} L_{\Sp} + \frac{1}{3 \sqrt{5}} \frac{\la{7} \la{11}} {\la{3}} \frac{M}{\sqrt{\MA \MS}} L_{16}\right) 
\end{gathered} 
\end{equation}
with normalization factors $N_u$ and $N_d$, where $L_x$, $\overline{L}_x$ denote the SM component with the quantum numbers of $h_d$, $h_d$ in the SO(10) representation $x$. 



After switching on $m$ the soft supersymmetry breaking terms and $\mu$-term are generated. The $\mu$-term is already present in the high energy Lagrangian and is of order $m$, the vev of $D_X$ generates sfermion and Higgs masses of order $m^2$ and the vev of $F_X$ gives rise to a $B_\mu$ term of order $m^2$. The heavy fields $5^{10}_i$ and $\overline{5}^{16}_i$ act as messengers of SUSY breaking to the gauginos who get masses of order $m^2/(16 \pi^2)$. The Goldstino will be mainly the fermion in $\Sip$ but gets also small contributions from the gaugino corresponding to the $U(1)_X$ generator and the fermion in $Z$. The corresponding scalar will get a mass of order $m^2$. 

%



\end{document}